\documentclass[
 reprint,
superscriptaddress,
 amsmath,amssymb,
 aps,
 pre,
]{revtex4-2}

\usepackage{placeins}
\usepackage{tikz}
\usepackage{subfigure}
\usepackage{placeins}
\usepackage{pgfplots}
 \pgfplotsset{compat=1.17}
 
\usetikzlibrary{decorations.pathreplacing}
\usepackage{graphicx}
\usepackage{dcolumn}
\usepackage{bm}
\usepackage{hyperref}

\begin{document}

\preprint{APS/123-QED}

\title{Master stability functions for metacommunities with two types of habitats}

\author{Alexander Krau\ss} \email{alexander.krauss@pkm.tu-darmstadt.de}
\affiliation{Technical University of Darmstadt, Institute for Condensed Matter Physics,  Hochschulstr. 6, 64289 Darmstadt}

\author{Thilo Gross} \email{thilo.gross@hifmb.de}
\affiliation{Helmholtz Institute for Functional Marine Biodiversity, University of Oldenburg (HIFMB), Ammerl\"ander Heerstra\ss e 231, 26129 Oldenburg.}
\affiliation{Alfred-Wegener-Institute for Marine and Polar Research, Bremerhaven, Germany, Am Handelshafen 12, 27570 Bremerhaven.}
\affiliation{University of Oldenburg, Oldenburg, Germany, Ammerl\"ander Heerstra\ss e 114-118, 26129 Oldenburg.}

\author{Barbara Drossel} \email{drossel@pkm.tu-darmstadt.de}
\affiliation{Technical University of Darmstadt, Institute for Condensed Matter Physics,  Hochschulstr. 6, 64289 Darmstadt}

\date{\today}

\begin{abstract}
\noindent{}Current questions in ecology revolve around instabilities in the dynamics on spatial networks and particularly the effect of node heterogeneity. We extend the master stability function formalism to inhomogeneous biregular networks having two types of spatial nodes. Notably, this class of systems also allows the investigation of certain types of dynamics on higher-order networks. Combined with the generalized modeling approach to study the linear stability of steady states, this is a powerful tool to numerically asses the stability of large ensembles of systems. We analyze the stability of ecological metacommunities with two distinct types of habitats analytically and numerically in order to identify several sets of conditions under which the dynamics can become stabilized by dispersal. Our analytical approach allows general insights into stabilizing and destabilizing effects in metapopulations. Specifically, we identify self-regulation and negative feedback loops between source and sink populations as stabilizing mechanisms and we show that maladaptive dispersal may be stable under certain conditions.
\end{abstract}

\maketitle

\section{\label{sec introduction}Introduction}

Understanding the factors that stabilize or destabilize nonlinear dynamics of networks is important when dealing with natural or human-made complex systems. 
Examples are the conservation of ecological networks \cite{gross2009, johnson2014trophic}, the analysis of regulatory networks in biological cells \cite{li2004yeast, davidich2008boolean}, the robust construction of electrical power grids \cite{rohden2012self, witthaut2012braess}, or flow optimization on human traffic networks \cite{gallotti2014anatomy}. Many of these systems can be conceptualized as complex multi-layer and/or higher-order networks. This allows for a richer and more detailed description of the system but complicates the analysis of the dynamics. For instance, in ecology plant-pollinator networks are coupled to herbivores that feed on the plants, and this can reduce the nestedness of the plant-pollinator network \cite{sauve2014structure}. The spatial coupling of chemical reaction networks in different biological cells or of local foodwebs on different habitat patches can give rise to Turing and wave instabilities that destroy steady states that would be stable in a spatially isolated network \cite{rohlf2009morphogenesis,brechtel2018}. Further, spatial coupling of a heterogeneous set of networks can stabilize networks that would be unstable otherwise, for instance in ecological source-sink systems \cite{holt1985,pulliam1988,dias1996,amarasekare2004,gravel2010}.

The theoretical study of multilayer systems is hampered by the computational costs that increase rapidly with the number of nodes and the number of parameters and possible network structures to be explored. An elegant tool to deal with the vast parameter space and the great variety of possible models is generalized modeling (GM) \cite{gross2006}, which focuses on linear stability of steady states and expresses the Jacobian of the system in terms of scale parameters and exponent parameters. Scale parameters quantify the relative contributions of the different types of gain and loss terms to the overall turnover of each species. Exponent parameters quantify how the different growth and loss terms change with the population densities in the vicinity of the considered steady state. By specifying the ranges of these parameters, a general class of models can be investigated without the need to precisely fix the functional forms of the growth and loss terms.

In networks with homogeneous nodes stability analysis can build on master stability functions (MSF). This idea is commonly credited to Ref. \cite{pecora1998master} who used it to study oscillator synchronization.  However, the mathematical formalism was already used in ecology by Othmer and Scriven \cite{othmer1971instability} in the context of pattern formation and was subsequently applied to complex networks by Segel and Levin \cite{segel1976application} and is also discussed in Refs. \cite{nakao2010turing,gibert2019laplacian}. 

The combination of GM and MSF was already used in Ref. \cite{brechtel2018} to study a multiplex metafoodweb \cite{gross2020modern}. These models describes a geographical network of $M$ habitat patches. Each of these patches is home to a complex food web of $N$ species, such that the dynamical dimension of the model is $NM$ and stability of steady states is captured by a Jacobian matrix of size $MN \times NM$. The GM+MSF approach disentangles the influence of the structure of the spatial network on stability from that of the local species network such that stability can be assessed by examining the eigenvalues of an $N \times N$ and an $M \times M$ matrix. Besides computational efficiency, the approach thus provides a deeper understanding of how the spatial and the ecological network structures impact stability. 

This MSF approach builds on the assumption that all nodes of the spatial network are equivalent, so that the species network has the same steady state on all habitats. Recently, this approach was extended to the case where only part of the species have populations on all habitats, while others are global species that have a single population that couples to its interaction partners on all habitats simultaneously \cite{brechtel2019far}. 

In this paper, we develop a generalization of the MSF approach to biregular heterogeneous systems, namely to metanetworks with different types of spatial nodes. These systems are regular graphs in the sense that every node of a given type is connected to the same number of neighboring nodes. Moreover, they are bipartite such that a node of a given type only interacts with the nodes of the respectively other type. 

Our motivation for choosing this particular system is twofold. First, from a purely operational perspective it provides a framework in which analytical insights can be gained which allows a deeper understanding of the important ecological question of node heterogeneity. Second, we are motivated by recent work on dynamics on hypergraph or higher-order dynamical networks \cite{ghorbanchian2021higherorder,mulas2020coupled}, which under certain conditions map to the biregular system class.  

In the ecological context the biregular system provides an adequate model for the dispersal of individuals between patches. When dispersing many higher animals enter a roaming-state where they are in transit between patches while looking for a new home range. In this roaming-state population dynamics (typically dominated by losses) can still occur such that there is population dynamics both on the links and the nodes of the geographical network, similary to the model in Ref. \cite{ghorbanchian2021higherorder}. Representing the nodes in a regular graph by one node type and the links connecting them as another node type leads to a biregular network of the type considered here.   

These ecological systems belong to the wider class of source-sink systems, which are ubiquitous in heterogeneous landscapes. Sources are high-quality habitats where populations exhibit a positive net growth rate, while sinks are poor-quality habitats with negative net growth rates. Coupling the two types of habitats can maintain populations in sinks that would otherwise go extinct, and outflow from sources as well as inflow from sinks can affect the stability of sources.  

Dispersal between sources and sinks cannot only happen through passive dispersal, but also through adaptive dispersal, for instance from overcrowded sources to lower-quality but less-crowded habitats \cite{pulliam1988,diffendorfer1998}. Such active dispersal can, however, also be maladaptive \cite{remevs2000,delibes2001}. A particular risk to wildlife are perceptual and ecological traps, where individuals disperse actively out of high-quality habitats or into low-quality habitats and thereby reduce their rate of reproduction \cite{robertson2006}. Many of these traps are due to recent changes caused by human interference \cite{fletcher2012}, such as reflecting artificial surfaces which attract mating water insects \cite{szaz2015,egri2017}, or developed mountain valleys that attract grizzly bears \cite{lamb2017}. Albeit detrimental to the population size, traps might sometimes help to stabilize population dynamics \cite{hale2015}. 
As a specific application of the methodological development presented in this paper, we will show that such stabilizing effects of perceptual and ecological traps are indeed possible under generic conditions.

In the following, we first develop the generalized MSF formalism for metanetworks with two types of spatial nodes. Then, we define a class of generalized ecological models of which the stability is explored with this method. We find that there are two types of dispersal turnover rates in source-sink dynamics, and specify general conditions under which increasing dispersal rates can have a stabilizing effect. Analytical calculations for one species in source-sink systems are supplemented by numerical evaluations of systems with foodwebs consisting of several species, giving similar stability conditions. Among these conditions, we will identify a subset that demonstrates a stabilizing effect of traps.

\section{\label{sec MSF}Diffusion-driven instabilities for two types of patches} 

We consider a system of $N$ species and $M$ patches. The dynamics for the population density $X_i^k$ of species $i$ on patch $k$ of a metanetwork such as an ecological metacommunity has the general form
\begin{eqnarray}
    \dot{X}_i^k = &&G_i^k(\bm{X}^k) - M_i^k(\bm{X}^k) \nonumber\\
                  &&+ \sum_{l=1}^M E_i^{kl}(\bm{X}^k,\bm{X}^l) - \sum_{l=1}^M E_i^{lk}(\bm{X}^l,\bm{X}^k)\, . \label{general model}
\end{eqnarray}
The first term describes local growth in patch $k$ due to primary production (for plants) and consumption of other species, the second local losses (``mortality'') due to predation and other causes of death, and the last two terms describe dispersal into and out of patch $k$. 
We assume that there are two types of patches, which we will call sources and sinks, with species in sources having a positive net growth rate $G_i^k(\bm{X}^k) - M_i^k(\bm{X}^k)$ and species in sinks having a negative one. We further assume that all sources have identical parameters (including patch size) and all sinks have identical parameters. The numbers of sources ($M^+$) and of sinks ($M^-$) will in general be different. We furthermore assume that sources are connected only to sinks and vice versa. In order to obtain identical steady states for all patches of the same type, they must have the same degree, thus the network has to be biregular. This means that at a steady state the population densities of all source patches are identical, $(\bm{X}^{k_+})^* = \bm{X}^+$ for $ k_+ \in \{1, \dots, M^+\}$ and those of all sinks are identical,    $(\bm{X}^{k_-})^* = \bm{X}^-$ for $ k_- \in \{M^+ + 1, \dots M^+ + M^-\}$. 

The stability of a steady state is determined by the Jacobian
\begin{equation}\label{jacobian}
\bm{J}_{(i+kN) (j+mN)} = \frac{\partial \dot{X}^k_i(\bm{X})}{\partial X^m_j}\bigg|_{\substack{\bm{X} = \bm{X}^*}}\; .
\end{equation}
When we separate the within-patch and between-patch terms, $\bm{J}$ has entries of the form
\begin{eqnarray}
\frac{\partial \dot{X}_i^k}{\partial X_j^k} &=& \bm{P}^k_{ij} - \sum_{l=1}^M \bm{C}^{kl}_{ij} \nonumber \\
\frac{\partial \dot{X}_i^k}{\partial X_j^l} &=& \hat{\bm{C}}^{kl}_{ij}
\end{eqnarray}
with the matrix elements that result from taking the derivatives of $G_i^k$ and $M_i^k$ in \eqref{general model}  being included in $\bm{P}^k$, and those that involve the dispersal terms $E_i^{kl}$ entering the matrices $\bm{C}^{kl}$ and $\hat{\bm{C}}^{kl}$.
Since the three types of matrices are identical for all patches of the same type, we can express the Jacobian in the compact block form
\begin{equation}
\label{jacobian biregular}
\bm{J} = 
 \begin{pmatrix}
  \bm{I}^+ \otimes (\bm{P}^+ - d^+ \bm{C}^+) & \bm{M} \otimes \hat{\bm{C}}^+ \\
  \bm{M}^T \otimes \hat{\bm{C}}^- & \bm{I}^- \otimes (\bm{P}^- - d^- \bm{C}^-)
 \end{pmatrix}\; ,
\end{equation}
where $\bm{I}^+$ ($\bm{I}^-$) is the identity matrix of dimension $M^+$ ($M^-$), and the matrix $\bm{M}$ denotes which patches are connected. It has $M^+$ rows and $M^-$ columns. We allow multiple links between the same pair of patches, and hence the entries of $\bm{M}$ are either $0$ or a natural number. Since all patches of one type have the same degree, $\bm{M}$ has a constant row sum equal to the source's degree $d^+$ and a constant column sum equal to the sink's degree $d^-$. The symbol $\otimes$ denotes the Kronecker product.

The eigenvalues $\lambda$ of the Jacobian are obtained by solving
\begin{equation}
\label{eigenvalue equation}
 \bm{J} \bm{u} = \lambda \bm{u} \; .
\end{equation}
The specific form \eqref{jacobian biregular} enables us to reduce the degree of this equation similarly to what has been done with the MSF approach by Brechtel et al.~\cite{brechtel2018}. To this purpose, we make the ansatz for the eigenvectors
\begin{equation}
\label{ansatz eigenvectors}
\bm{u} = 
 \begin{pmatrix}
  \bm{w}^+ \otimes \bm{v}^+ \\
  \bm{w}^- \otimes \bm{v}^- \\
 \end{pmatrix} \, .
\end{equation}
The $N$-dimensional normalized vectors $\bm{v}^+$ and $\bm{v}^-$ give the relative contribution of the different species to the eigenvectors within sources and sinks respectively, and $\bm{w}^+$ and $\bm{w}^-$ denote the weights of these eigenvectors on the different source and sink patches. 

In the following, we will show that we can find all solutions of the eigenvalue Eq. \eqref{jacobian biregular} by specifying that
\begin{equation}\bm{M}^T \bm{w}^+= \beta \bm{w}^- \quad \text{and}\quad \bm{M}\bm{w}^-=\alpha \bm{w}^+\, . \label{alphabeta}
\end{equation}
Inserting this together with the ansatz \eqref{ansatz eigenvectors} into the eigenvalue Eq. \eqref{eigenvalue equation} leads to a reduced eigenvalue equation with a reduced Jacobian  $\bm{j}$,
\begin{eqnarray}
 \bm{j} \begin{pmatrix}
  \bm{v}^+ \\
  \bm{v}^-
 \end{pmatrix}
 &\equiv&  \begin{pmatrix}
\bm{P}^+ - d^+ \bm{C}^+ & \alpha \hat{\bm{C}}^+ \\
  \beta  \hat{\bm{C}}^-   & \bm{P}^- - d^- \bm{C}^-
 \end{pmatrix}
 \begin{pmatrix}
  \bm{v}^+ \\
  \bm{v}^-
 \end{pmatrix} \nonumber\\
&=& \lambda
 \begin{pmatrix}
  \bm{v}^+ \\
  \bm{v}^-
 \end{pmatrix}\; , \label{reduced eigenvalue equation}
\end{eqnarray}
provided that $\bm{w}^+$ and $\bm{w}^-$ are nonvanishing. This reduced Jacobian is of dimension $2N$ instead of $M N$, while the influence of topology is captured in the coefficients $\alpha$ and $\beta$. Since the reduced eigenvalue Eq. \eqref{reduced eigenvalue equation} depends only on the product $\alpha \beta$, we can choose
\begin{equation}
 \alpha = \beta
\end{equation}
for the purpose of calculating $\lambda$ without loss of generality. One class of solutions for the eigenvalue Eq. \eqref{eigenvalue equation} can be obtained by transforming \eqref{alphabeta} to 
\begin{equation}
\label{singular equation 1}
 \beta^2 \bm{w}^+ = \bm{M} \bm{M}^T \bm{w}^+
\end{equation}
or 
\begin{equation}
\label{singular equation 2}
 \beta^2 \bm{w}^- = \bm{M}^T \bm{M} \bm{w}^-\; .
\end{equation}
Thus the positive $\beta$ are the singular values \cite{paige1981} of $\bm{M}$, i.e., ${\beta^2}$ are the joint eigenvalues of $\bm{M} \bm{M}^T$ and $\bm{M}^T \bm{M}$. Their quantity is $\min\{M^+,M^-\}$. The  largest of these singular values is $\sqrt{d^+ d^-}$ because $\bm{M}^T \bm{M}$ and $\bm{M} \bm{M}^T$ have a constant row sum of $d^+ d^-$. Since $\bm{M}^T \bm{M}$ and $\bm{M} \bm{M}^T$ are symmetric and positive semidefinite,  all of their eigenvalues are real and non-negative, i.e., $0 \leq \beta^2 \leq d^+ d^-$. From this, we find $2N * \min\{M^+,M^-\}$ eigenvalues of the full Jacobian $\bm{J}$ in Eq. \eqref{jacobian biregular} by solving the reduced eigenvalue Eq. \eqref{reduced eigenvalue equation} for each singular value $\beta \in S$, with $S$ being the set containing all singular values of $\bm{M}$. 

The remaining solutions are related to eigenvalues $\alpha\beta=0$ of $\bm{M} \bm{M}^T$ or $\bm{M}^T \bm{M}$ ($\alpha \neq \beta$), whichever of these two matrices has the larger dimension. This matrix has $|M^+-M^-|$ eigenvalues $0$ in addition to the singular values of $\bm{M}$ (among which there might also be $0$s). We find the solutions associated with this eigenvalue by setting  $\alpha=0$ and $\bm{w}^-=0$ in \eqref{alphabeta} if $M^+>M^-$, and $\beta = 0$ and $\bm{w}^+=0$  otherwise. In this case the eigenvalue Eq. \eqref{eigenvalue equation} simplifies to the two equations

\begin{subequations}
\label{jreduced}
\begin{eqnarray}
\bm{M}^k\bm{w}^k &=&0\\
    (\bm{P}^k - d^k \bm{C}^k)\bm{v^k} &=& \lambda \bm{v^k}
\end{eqnarray}
\end{subequations}

where $k=+$ and $\bm{M}^+= \bm{M}^T$ if $M^+>M^-$ and $k=-$ and $\bm{M}^-= \bm{M}$  if $M^->M^+$. 
The corresponding modes are nonvanishing only on the more numerous type of patches. This gives $N|M^+-M^-|$ additional solutions of Eq. \eqref{eigenvalue equation}. Altogether we thus have found all $2N * \min\{M^+,M^-\} + N*|M^+-M^-| = NM$ solutions of the full eigenvalue equation for $\bm{J}$. Compared to the direct calculation of the eigenvalues of $\bm{J}$, this procedure saves a significant amount of computation time when the system has many patches. 

The largest singular value $\beta = \sqrt{d^+ d^-}$ is obtained when $\bm{w}^+$ and $\bm{w}^-$ are vectors with all entries being 1. In this case, the dynamics close to the steady state are equivalent for all source and sink patches, respectively, and the associated eigenvalues are those of a system with one source and one sink patch. Since all eigenvalues $\lambda$ must be negative for a stable steady state, the steady state of the $M$-patch system can only be stable if the corresponding two-patch system is stable. 

However, stability of the two-patch system is not sufficient for stability of the entire network, as heterogeneous modes, which have different amplitudes on different source and/or sink patches, may also be unstable, leading to a generalized version of diffusion-driven Turing instabilities or of wave instabilities.

\section{\label{sec GM}Generalized modeling of source-sink dynamics}

In order to evaluate the stability of source-sink metacommunities, we use the GM approach \cite{gross2006}. We will choose a form for Eq. \eqref{general model} that naturally reflects the differences between sources and sinks. Local production is larger than local mortality in sources and vice versa in sinks. Similarly, emigration is larger than immigration in sources and smaller than immigration in sinks. The steady state is defined by $\dot{X}_i^k = 0$. Then the difference between growth and mortality is equal to the difference between immigration and emigration in both patch types. We normalize all population densities and functions to their values at the considered steady state (marked with $*$) and use lower-case symbols for these normalized quantities, setting ${x}_i^k = {X}_i^k/({X}_i^k)^*$ and $g_i^k(\bm{x}^k) = G_i^k(\bm{X}^k)/[G_i^k(\bm{X}^k)]^*$ and similarly for the other three functions. Eq. \eqref{general model} then can be put into a form that reflects directly the surplus growth and emigration of sources and the surplus immigration and mortality of sinks (see Appendix \ref{Appendix_GM} for a detailed calculation):

\begin{eqnarray}
\dot{x}_i^{k_+} =& \alpha_{P_i}^{+} &\bigg[g_i^{k_+}(\bm{x}^{k_+})  - m_i^{k_+}(\bm{x}^{k_+}) \bigg] \nonumber\\
	    &+ \alpha_{S_i}^{+} &\bigg[g_i^{k_+}(\bm{x}^{k_+}) - \frac 1 {d^+}\sum_{k_-} e_i^{{k_-}{k_+}}(\bm{x}^{k_-}, \bm{x}^{k_+})\bigg] \nonumber\\
	    &+ \alpha_{C_i}^{+} &\bigg[\frac 1 {d^+}\sum_{k_-}  e_i^{{k_+}{k_-}}(\bm{x}^{k_+}, \bm{x}^{k_-}) \nonumber\\
	    &&- \frac 1 {d^+}\sum_{k_-}  e_i^{{k_-}{k_+}}(\bm{x}^{k_-}, \bm{x}^{k_+})\bigg] \label{normalized source}
\end{eqnarray}
for all sources and 
\begin{eqnarray}
\dot{x}_i^{k_-} =& \alpha_{P_i}^{-} &\bigg[g_i^{k_-}(\bm{x}^{k_-})  - m_i^{k_-}(\bm{x}^{k_-}) \bigg] \nonumber\\
	    &+ q\alpha_{S_i}^{+} &\bigg[\frac 1{d^-}\sum_{k_+}  e_i^{{k_-}{k_+}}(\bm{x}^{k_-}, \bm{x}^{k_+}) - m_i^{k_-}(\bm{x}^{k_-}) \bigg] \nonumber\\
	    &+ q\alpha_{C_i}^{+} &\bigg[\frac 1{d^-}\sum_{k_+}  e_i^{{k_-}{k_+}}(\bm{x}^{k_-}, \bm{x}^{k_+}) \nonumber\\
	    &&- \frac {1}{d^-}\sum_{k_+} e_i^{{k_+}{k_-}}(\bm{x}^{k_+}, \bm{x}^{k_-})\bigg] \label{normalized sink}
\end{eqnarray}
for all sinks. The $d^k$ denote the degree of patch $k$, and each of the sums has exactly $d^k$ nonzero terms. In general, we assume that dispersal depends on a species' own population density (and on the density of other species such as a predator or prey) on the donor and target patch.

The per-capita biomass turnover rates $\alpha$ have an intuitive meaning that will be useful for the stability analysis further below. The rate $\alpha_{P_i}^k$ captures the turnover due to local dynamics. In sources $\alpha_{P_i}^{+}$ is equal to the mortality rate and in sinks $\alpha_{P_i}^{-}$ is equal to the growth rate. Turnover due to pure dispersal dynamics is denoted with $\alpha_{C_i}^{+}$ being the rate of immigration to sources and $q \alpha_{C_i}^{+}$ the rate of emigration from sinks. We therefore call them dispersal turnover. These are proportional to each other with the factor $q$ because the total biomass leaving a donor patch needs to be equal to the biomass arriving at the target patch (see Appendix \ref{Appendix_GM} for further information). In addition to these two rates, there is an excess growth (not compensated by mortality) and an excess emigration (not compensated by immigration) in sources, and an excess mortality and immigration in sinks. This excess leads to a net source-sink flow rate $\alpha_{S_i}^+$ in sources and $q\alpha_{S_i}^+$ in sinks. It is a direct measure of the strength of source-sink dynamics, which are absent if $\alpha_{S_i}^+ = 0$. We have thus two distinct types of dispersal rates in source-sink dynamics. The rate $\alpha_{C_i}^+$ is due to dispersal between sinks and sources, while $\alpha_{S_i}^+$ is the rate of the net flow from sources to sinks driven by excess growth in sources. Below, we will investigate the impact of these two scale parameters on stability.

The stability of the system is evaluated by analyzing  the Jacobian of Eqs. \eqref{normalized source} and \eqref{normalized sink},
\begin{equation}\label{jacobian2}
\bm{J}_{(i+kN) (j+mN)} = \frac{\partial \dot{x}^k_i(\bm{x})}{\partial x^m_j}\bigg|_{\substack{\bm{x} = \bm{x}^*}}\; .
\end{equation}
It has the same eigenvalues as the Jacobian \eqref{jacobian} and can be evaluated for our biregular system with the methods described in Sec. \ref{sec MSF}.

For evaluating the Jacobian we need the derivatives of the normalized functions at the steady state, which are the so-called exponent parameters. In our model, we have the following exponent parameters:
\begin{eqnarray}
\label{exponent parameters}
&\phi_i^k &= \frac{\partial}{\partial x_i^k} g_i^k (x_i^k) \bigg|_{\substack{\bm{x} = \bm{x}^*}}\; , \quad
\mu_i^k = \frac{\partial}{\partial x_i^k} m_i^k (x_i^k) \bigg|_{\substack{\bm{x} = \bm{x}^*}}\; , \nonumber\\
&\phi_{ij}^k &= \frac{\partial}{\partial x_j^k} g_i^k (x_i^k) \bigg|_{\substack{\bm{x} = \bm{x}^*}} \; , \quad
\mu_{ij}^k = \frac{\partial}{\partial x_j^k} m_i^k (x_i^k) \bigg|_{\substack{\bm{x} = \bm{x}^*}} \; , \nonumber\\
&\hat{\omega}_i^{kl} &= \frac{\partial}{\partial x_i^k} e_i^{kl} (\bm{x}^k, \bm{x}^l) \bigg|_{\substack{\bm{x} = \bm{x}^*}}\; , \nonumber\\
&\omega_i^{kl} &= \frac{\partial}{\partial x_i^l} e_i^{kl} (\bm{x}^k, \bm{x}^l) \bigg|_{\substack{\bm{x} = \bm{x}^*}}\; , \nonumber\\
&\hat{\kappa}_{ij}^{kl} &= \frac{\partial}{\partial x_j^k} e_i^{kl} (\bm{x}^k, \bm{x}^l) \bigg|_{\substack{\bm{x} = \bm{x}^*}} \; , \nonumber\\
&\kappa_{ij}^{kl} &= \frac{\partial}{\partial x_j^l} e_i^{kl} (\bm{x}^k, \bm{x}^l) \bigg|_{\substack{\bm{x} = \bm{x}^*}} 
\end{eqnarray}
with $i \neq j$. 

The exponent parameters are logarithmic derivatives of the unnormalized functions with respect to the population density (also called elasticities). They give the power law exponent that characterizes the relationship between the function and the population density in the vicinity of the steady state. For instance, an exponent parameter of $1$ describes a linear relationship, an exponent of $2$ a quadratic relationship and an exponent of $-1$ an inversely proportional relationship between a function and its argument. Table \ref{table generalized parameters} summarizes all generalized parameters and their ecological interpretation.

Ecological considerations fix the sign and the realistic range of values of exponent parameters. In particular, $\phi_i^k$ and $\mu_i^k$ must be positive since we expect local growth and mortality to increase with population density. This increase is expected to be linear ($\phi_i^k=1$, $\mu_i^k=1$) for simple Lotka Volterra models. Predator saturation and finite handling times of prey typically make growth sublinear ($\phi_i^k<1$). Intra-specific competition (e.g., due to limitations of space, nutrients or nesting sites) leads to self-regulation, making mortality superlinear ($\mu_i^k>1$). This induces a self-regulating negative feedback between net growth rate and population density which is stabilizing \cite{barabas2017self}. The opposite situation is also possible. A population may be subject to strong intra-specific facilitation, also known as the Allee effect, which typically occurs at lower population densities, e.g., due to a better protection against predators in large groups or an increased probability to find mating partners. In this situation we expect growth to correlate with population density in a superlinear ($\phi_i^k>1$) and mortality to correlate in a sublinear ($\mu_i^k<1$) manner. This is known to have a destabilizing effect.

Movement behavior can be described by how dispersal rates depend on population densities in the donor ($\omega_i^{kl}$) or target ($\hat{\omega}_i^{kl}$) patch. If dispersal is passive, then we expect a fixed  proportion of the population in the donor patch to emigrate. Then the per-capita emigration rates are constant. In this case dispersal rates are independent of the density in the target patch $k$ and linear in the density of the donor patch $l$ such that $\omega_i^{kl}=1$ and $\hat{\omega}_i^{kl}=0$. Individuals might be more likely to avoid their own species ($\omega_i^{kl}>1$, $\hat{\omega}_i^{kl}<0$, e.g., due to competition) or other species ($\kappa_{ij}^{kl}>0$, $\hat{\kappa}_{ij}^{kl}<0$, e.g., due to predation by that species) or they might be more likely to seek their own species ($\omega_i^{kl}<1$, $\hat{\omega}_i^{kl}>0$, e.g., due to an Allee effect) or other species ($\kappa_{ij}^{kl}<0$, $\hat{\kappa}_{ij}^{kl}>0$, e.g., when seeking prey). In cases of strong facilitation, emigration from a patch might even be negatively correlated with its density ($\omega_i^{kl}<0$).
Movement behavior might even lead to a decrease of per-capita growth rates. Such movement is maladaptive and can give rise to so-called traps. We will show below that such maladaptive situations may arise under generic conditions.

These considerations show that by imposing conditions on the generalized parameters one can specify classes of ecological systems that share essential characteristics of their dynamics. In Table \ref{table generalized parameters} we list the realistic ranges in which the generalized parameters can lie. In the next section, we will show that there are qualitatively different classes of systems within these ranges that show different stability properties.

\section{\label{sec results}Analytical results for one species on multiple patches}

The reduced eigenvalue Eq. \eqref{reduced eigenvalue equation} becomes analytically solvable if only one species is present. This analytical solution provides a wealth of insights about the stability of the system, part of which we will derive in the following. 

\subsection{\label{sec EV}Eigenvalues of the Jacobian}

As outlined in Sec. \ref{sec MSF}, there are two types of eigenvalues of the Jacobian. First, there are two eigenvalues $\lambda_i(\beta)$ ($i \in {1, 2}$) for each singular value $\alpha=\beta$ of the matrix $\bm{M}$. These eigenvalues are obtained by solving the reduced eigenvalue Eq. \eqref{reduced eigenvalue equation}. For one species, the reduced Jacobian $\bm{j}$ simplifies to
\begin{equation}
\label{reduced jacobian}
\bm{j} = 
 \begin{pmatrix}
  P^+ - d^+ C^+ & \beta \hat{C}^+ \\
  \beta \hat{C}^- & P^- - d^- C^-
 \end{pmatrix}
\end{equation}
with
\begin{eqnarray}
 &P^+ &= \alpha_P^+ (\phi^+ - \mu^+) + \alpha_S^+ \phi^+ \label{pplus} \\
 &P^- &= \alpha_P^- (\phi^- - \mu^-) - q \alpha_S^+ \mu^- \label{pminus} \\
 &C^+ &= [\alpha_C^+ (\omega^{-+} - \hat{\omega}^{+-}) + \alpha_S^+ \omega^{-+}]/d^+ \label{cplus} \\
 &C^- &= [q\alpha_C^+ (\omega^{+-} - \hat{\omega}^{-+}) - q \alpha_S^+ \hat{\omega}^{-+}]/d^-  \\
 &\hat{C}^+ &= [\alpha_C^+ (\omega^{+-} - \hat{\omega}^{-+}) - \alpha_S^+ \hat{\omega}^{-+}] /d^+  \\
 &\hat{C}^- &= [q\alpha_C^+ (\omega^{-+} - \hat{\omega}^{+-}) + q \alpha_S^+ \omega^{-+}]/d^-\; .
\end{eqnarray}

Second, if the number of source patches $M^+$ is different from the number of sink patches $M^-$, then there is one additional eigenvalue $\lambda_0^k$ that is obtained by setting $\alpha=0$ ($\beta=0$) if $M^+> M^-$ ($M^+< M^-$) and solving Eq.~\eqref{jreduced}. The considered steady state is stable if the real part of all eigenvalues of the reduced Jacobian $\bm{j}$ and the additional eigenvalue $\lambda_0^k$ (which is real) are negative, i.e., if $\text{Re}[\lambda_i(\beta)] < 0 \quad \forall \beta \in S$ and if
  \begin{equation}\label{lambda add}
     P^k-d^kC^k=\lambda_0^k < 0\; ,
 \end{equation}
 with $k=+$ if $M^+>M^-$ and $k=-$ if $M^->M^+$. The biregular spatial distribution of patches affects stability only through the parameters $d^+$, $d^-$, and $\beta$.

In the absence of dispersal, the dispersal-dependent turnover rates $\alpha_S^+$ and $\alpha_C^+$ vanish. Then the Jacobian has a diagonal form with the entries $P^+$ and $P^-$.  According to eqs.~\eqref{pplus}  and \eqref{pminus}, both types of patches are stable if \begin{equation}
\label{result local}
    \phi^+<\mu^+\; , \quad\quad \phi^-<\mu^-\; .
\end{equation}

In the following we refer to the term
\begin{equation}
\label{local feedback}
    \alpha_P^k(\phi^k-\mu^k)
\end{equation}
as local feedback. If the local feedback in both patch types is negative, then local dynamics exhibit a stable steady without dispersal because the local populations are able to self-regulate their sizes.

\subsection{\label{sec NSSF}When is the net source sink flow stabilizing?}

We consider systems that are unstable in the absence of dispersal and explore under which conditions an increase of the net source-sink flow $\alpha_S^+$ can stabilize the steady state (see Appendix \ref{Appendix_NSSF} for the complete calculation).

For the sake of simplicity we assume that $\alpha_C^+=0$. We identify two effects of $\alpha_S^+$ on the entries of the Jacobian. 

First, the additional eigenvalue $\lambda_0^k$ \eqref{lambda add} and the diagonal entries $P^k-d^kC^k$ of the Jacobian depend linearly on $\alpha_S^+$.  
We call the diagonal entries intrapatch feedbacks since they indicate how the net growth rate of a population correlates with its local size at the steady state. Negative intrapatch feedbacks at a steady state are known to increase stability \cite{gross2009, barabas2017self} because they indicate that a population is able to self-regulate its population size. Due to dispersal, these intrapatch feebacks are shifted by 
\begin{equation}
\label{intrapatch feedback S}
\alpha_S^+(\phi^+-\omega^{-+}) \quad\text{and}\quad q\alpha_S^+(\hat{\omega}^{-+}-\mu^-)
\end{equation}
for sources and sinks respectively. We say that these terms are \emph{induced} by the net source-sink flow $\alpha_S^+$. We therefore call these terms "induced intrapatch feedback".

Second, the off-diagonal entries of the reduced Jacobian become nonzero due to dispersal. The product 
\begin{equation}
\label{feedback loop}
    - q (\alpha_S^+)^2 \frac{\beta^2}{d^+d^-} \hat{\omega}^{-+} \omega^{-+}
\end{equation}
of the off-diagonal elements of $\bm{j}$ indicates a feedback \cite{neutel2014interaction} between the source and sink populations. We call this "induced interpatch feedback".
Whether this feedback is stabilizing or destabilizing, depends on the sign of $- \hat{\omega}^{-+}\omega^{-+}$. For example, if $\omega^{-+}>0$, then a positive perturbation to source populations increases immigration to sinks and thus the population size of sinks. But because of $\hat{\omega}^{-+}<0$ this also decreases emigration from sources and thus increases the source population further. If this positive interpatch feedback is not opposed by sufficiently strong negative intrapatch feedbacks \eqref{intrapatch feedback S}, then a positive feedback loop is present which destabilizes the equilibrium. By the same type of reasoning, the case $-\hat{\omega}^{-+}\omega^{-+}<0$ implies a negative interpatch feedback, which has a stabilizing effect. 

Since all Jacobian eigenvalues need to be negative for all $\beta^2$, it is sufficient to restrict the analysis to values for $\beta^2$ where \eqref{feedback loop} is maximal (see Appendix \ref{Appendix_NSSF}), which is $\beta^2 = d^+d^-$ if $- \hat{\omega}^{-+}\omega^{-+}>0$ or $\beta^2=\beta_0^2=\text{min}(\beta \in S)^2$ if $- \hat{\omega}^{-+}\omega^{-+}<0$. In the latter case, stability is more easily achieved when  $\beta_0$ is larger. This means that the topology of the source-sink network affects stability, since the $\beta^2$ are the eigenvalues of $\bm{M} \bm{M}^T$ and $\bm{M}^T \bm{M}$, see \eqref{singular equation 2}. 

In our analytical considerations, we set $\beta_0^2=0$ in order to keep calculations feasible, which means that our stability criteria are sufficient, but that part of the systems that do not satisfy the criteria can also be stable.

We obtain five different ways how a net source-sink flow can stabilize a system that would be unstable in the absence of dispersal. The first three cases are such that there exists a threshold value of $\alpha_S^+$ above which the maximal real part of all eigenvalues is negative. This means that the system eventually becomes stable when $\alpha_S^+$ is increased far enough. A prerequisite for this to be possible is that dispersal and density are positively correlated in sources ($\omega_i^{-+}>0$). We find (for further details see Appendix \ref{Appendix_LNSSF}):

\begin{quote}
\textit{I A sufficiently large net source-sink flow $\alpha_S^+$ stabilizes source-sink dynamics if it induces a negative intrapatch feedback in all patches ($\phi^+<\omega^{-+}$ and $\hat{\omega}^{-+}<\mu^-$) and no or a negative interpatch feedback ($-\omega^{-+}\hat{\omega}^{-+}\leq 0$).}
    \end{quote}

\begin{quote}
\textit{II If $\alpha_S^+$ induces a positive interpatch feedback ($-\omega^{-+}\hat{\omega}^{-+}>0$), a stabilization of source-sink dynamics is possible if it also induces a negative intrapatch feedback in all patches ($\phi^+<\omega^{-+}$ and $\hat{\omega}^{-+}<\mu^-$), which has to be strong enough to suppress a potential positive feedback loop between source and sink populations. This is the case if $|\hat{\omega}^{-+}| < {\mu^-} \left(\frac{\omega^{-+}}{\phi^+}-1\right)$.}
    \end{quote}

\begin{quote}
\textit{III A sufficiently large net source-sink flow $\alpha_S^+$ stabilizes source-sink dynamics even if it induces a positive intrapatch feedback in one of the two patch types ($\phi^+>\omega^{-+}$ or $\hat{\omega}^{-+}>\mu^-$). This has to be in the less numerous patch type (so that condition \eqref{lambda add} does not apply to it), and the absolute value of the induced feedback has to be smaller than the negative intrapatch feedback which is induced in the more numerous patch type. In addition, $\alpha_S^+$ must induce a sufficiently strong negative feedback loop between source and sink populations ($-\omega^{-+}\hat{\omega}^{-+}<0$, $\beta_0^2>0$) to overcome local Allee effects (see Appendix \ref{Appendix_LNSSF}).}
    \end{quote}

In the remaining two cases the model is unstable not only in the absence of dispersal but also in the limit of very large net source-sink flow $\alpha_S^+$. This means that there exists an intermediate interval of $\alpha_S^+$ values for which the system is stable. In order to specify these two cases, it is useful to define the resistances $R_S^+$ and $R_S^-$ of sources and sinks to changes of the net-source sink flow $\alpha_S^+$,
\begin{equation}\label{resistance}
    R_S^+ = -\alpha_P^+  \frac{\phi^+ - \mu^+}{\phi^+-\omega^{-+}} \, , \quad R_S^- = -\frac{\alpha_P^-}{q}  \frac{\phi^- - \mu^-}{\hat{\omega}^{-+}-\mu^-} \, .
\end{equation}
Since a diagonal entry of the Jacobian \eqref{reduced jacobian} is zero if $\alpha_S^+=R_S^k$, these resistances also measure how sensitive the diagonal entries (intrapatch feedbacks) are to changes in $\alpha_S^+$.

We find the following two situations where the system can be stable for an intermediate range of $\alpha_S^+ $ values (for further details on the analysis and for the specification of necessary conditions see Appendix \ref{Appendix_INSSF}): 
\begin{quote}
\textit{IV If the resistance  of patch type $k_1$ with a negative local feedback ($\phi^{k_1}<\mu^{k_1}$) is larger than the resistance ($R_S^{k_1}>R_S^{k_2}$) of the other patch type $k_2$ with a positive local feedback ($\phi^{k_2}>\mu^{k_2}$), then the net source-sink flow $\alpha_S^+$ can stabilize the system for an intermediate range of values. This requires that $\alpha_S^+$ induces a positive intrapatch feedback \eqref{intrapatch feedback S} in the former ($k_1$) and a negative intrapatch feedback in the latter ($k_2$). The induced interpatch feedback may be negative or positive, but has to be sufficiently weak in the latter case.}
\end{quote}

 \begin{quote}
     \textit{
      V If the net source-sink flow $\alpha_S^+$ induces a stabilizing negative intrapatch feedback in both sources and sinks ($\phi^+ < \omega^{-+}$ and $\hat{\omega}^{-+} < \mu^-$), then an intermediate range of stability is possible if $\alpha_S^+$ also induces a destabilizing positive interpatch feedback ($-\hat{\omega}^{-+}\omega^{-+} > 0$), which has to dominate the induced negative intrapatch feedbacks for a large net source-sink flow, which is the case if $|\hat{\omega}^{-+}| > {\mu^-} \left(\frac{\omega^{-+}}{\phi^+}-1\right)$. It is required that one patch type has a negative local feedback ($\phi^{k_1}<\mu^{k_1}$), while the other has a positive one ($\phi^{k_2}>\mu^{k_2}$), with the resistances satisfying $-R_S^{k_1}>R_S^{k_2}$.}
 \end{quote}

\subsection{\label{sec DT}When is dispersal turnover stabilizing?}

Next, we analyze the conditions under which the dispersal turnover $\alpha_C^+$ stabilizes the system. Proceeding similarly as before, we identify the intrapatch feedbacks 
\begin{equation}
\label{intrapatch feedback C}
\alpha_C^+(\hat{\omega}^{+-}-\omega^{-+}) \quad\text{and}\quad q\alpha_C^+(\hat{\omega}^{-+}-\omega^{+-})
\end{equation}
which are induced by the dispersal turnover $\alpha_C^+$. Again we define the resistance of sources $R_C^+$ and resistance of sinks $R_C^-$ to changes in dispersal turnover $\alpha_C^+$,
\begin{eqnarray}
    R_C^+ &= -  \frac{\alpha_P^+(\phi^+ - \mu^+) +\alpha_S^+\phi^+}{\hat{\omega}^{+-}-\omega^{-+}} \, , \nonumber\\ \quad R_C^- &= -\frac{\alpha_P^-(\phi^- - \mu^-) -q\alpha_S^+\mu^-}{q(\hat{\omega}^{-+}-\omega^{+-})} \, .\label{resistance C}
\end{eqnarray}
Compared to the resistances to the net source-sink flow \eqref{resistance}, the denominators are replaced by the intrapatch feedbacks induced by the dispersal turnover \eqref{intrapatch feedback C}. The numerators
\begin{equation}\label{extendedlocalfeedback}
    \alpha_P^+(\phi^+-\mu^+)+\alpha_S^+\phi^+ \quad \text{ and } \quad 
    \alpha_P^-(\phi^- - \mu^-)-q\alpha_S^+\mu^-
\end{equation}
represent extended local feedbacks, which have an additional term due to source-sink dynamics. We identify two regimes of stability for $\alpha_C^+$, one for sufficiently large and one for intermediate $\alpha_C^+$. First, we find (see Appendix \ref{Appendix_LDT} for further details on the analysis):

\begin{quote}
\textit{
VI A sufficiently large dispersal turnover $\alpha_C^+$ is stabilizing if it induces a negative intrapatch feedback in both patch types ($\hat{\omega}^{+-}<\omega^{-+}$ and $\hat{\omega}^{-+}<\omega^{+-}$) and if their extended local feedbacks \eqref{extendedlocalfeedback} have opposite signs ($k_1$ with a negative and $k_2$ with a positive extended local feedback). Additionally, the absolute value of the resistance of the patch type with a negative extended local feedback has to be larger than the absolute value of the resistance of the other ($R_C^{k_1}>-R_C^{k_2}$).
}
\end{quote}

For the intermediate regime of stability we assume $\alpha_S^+=0$ for simplicity and find (see Appendix \ref{Appendix_IDT} for details):

\begin{quote}

\textit{VII The dispersal turnover $\alpha_C^+$ can stabilize the system for an intermediate range of values if the induced intrapatch feedback in a patch type $k_1$ with a negative local feedback ($\phi^{k_1}<\mu^{k_1}$) is positive ($\hat{\omega}^{k_1k_2}>\omega^{k_2k_1}$) and the induced intrapatch feedback in a patch type $k_2$ with a positive local feedback ($\phi^{k_2}>\mu^{k_2}$) is negative ($\hat{\omega}^{k_2k_1}<\omega^{k_1k_2}$). Further it is required that the resistance of the former is larger than the resistance of the latter ($R_C^{k_2} > R_C^{k_1}$).
}
\end{quote}
These conditions for stabilization do not depend on induced interpatch feedbacks. Hence stabilizing effects of dispersal turnover mainly depend on the intrapatch feedbacks it induces, though thresholds for stability may be shifted due to induced interpatch feedbacks.
\subsection{Maladaptive dispersal can be stabilizing}

The conditions identified in the previous sections are so general that they include instances of maladaptive dispersal. To model maladaptive dispersal we consider the exponent parameters that are related to the local per-capita net growth rates
\begin{equation}\label{eq:defW}
    W_i^k(\bm{X}^k) = \frac{G_i^k(\bm{X}^k) - M_i^k(\bm{X}^k)}{X_i^k}\, ,
\end{equation}
as these are a measure for patch quality. The exponent parameters related to these net growth rates can be expressed in terms of the already defined parameters as
\begin{eqnarray}
&\eta_i^+ =\frac{\partial W_i^{k_+}}{\partial X_i^{k_+}} \frac{X_i^{+*}}{W_i^{+*}} \bigg|_{\substack{\bm{X} = \bm{X}^*}} = \frac{\alpha_{P_i}^+}{\alpha_{S_i}^+} (\phi_i^+ - \mu_i^+) + \phi_i^+ -1 \; , \nonumber\\
&\eta_i^- = \frac{\partial W_i^{k_-}}{\partial X_i^{k_-}} \frac{X_i^{-*}}{W_i^{-*}} \bigg|_{\substack{\bm{X} = \bm{X}^*}} = \frac{\alpha_{P_i}^-}{\alpha_{S_i}^-} (\phi_i^- - \mu_i^-) - \mu_i^- -1 \quad
\end{eqnarray}
for sources and sinks respectively.

Usually it is expected that individuals disperse adaptively such that they increase their local per-capita net growth rate, and thus on average they should prefer sources over sinks. However, the local per-capita net growth rate can decrease with density ($\eta_i^+<0$) even if $W_i^k(\bm{X}^k)>0$, for instance if there is competition for a limited resource. Then it might be favorable for part of the population to move from a source to a sink. In addition to this adaptive type of movement, maladaptive movement is also possible. In this case the per-capita net growth rate decreases due to dispersal. 

Here we consider two cases of maladaptive movement. First, if a source is subject to an Allee effect or growth facilitation ($\eta_i^+>0$), then an increase in the source's population size also increases $W_i^k(\bm{X}^+)$. Then  movement is certainly maladaptive if individuals additionally are more likely to leave ($\omega_i^{-+}>1$, positive density dependence) or avoid to settle ($\hat{\omega}_i^{+-} < 0$) in the source when the source population becomes larger. This can be called a perceived trap, since individuals falsely identify a high-quality habitat as a bad one and avoid it \cite{gilroy2007,patten2010}. Second, if a sink is subject to competition ($\eta_i^-<0$), then an increase in its population size decreases $W_i^k(\bm{X}^-)$. Then movement is certainly maladaptive if individuals are more likely to settle ($\hat{\omega}_i^{-+} > 0$) or to stay ($\omega_i^{+-}<1$, negative density dependence) in sinks when the sink population size is larger. In this case we have an ecological trap since individuals actively prefer to settle in a poor-quality habitat \cite{battin2004}.

We find from Sec. \ref{sec NSSF} that there always exists a threshold value for $\alpha_S^+$ above which the system is stable irrespective of the sign of local feedbacks $\alpha_P^k(\phi^k-\mu^k)$, as long as the according conditions are satisfied. Hence we can choose $\eta^+>0$ ($\eta^-<0$) and $\omega^{-+}>1$ ($\hat{\omega}^{-+}>0$), so that there exist stable steady states according to cases I-III (I and III) if we chose the other parameters accordingly.

There also exist intermediate ranges of $\alpha_S^+$ for which stabilization is possible if a trap is present. For instance in case IV we can choose $\phi^+-\mu^+>0$ and $\omega^{-+}>\phi^+>1$ and obtain $\eta^+>0$, implying a perceptual trap. Then a stable steady state is obtained when the other parameters are chosen accordingly. Similarly we can choose $\phi^+-\mu^+<0$ and $\hat{\omega}^{-+}>\mu^->0$ with $\eta^-<0$ and obtain an ecological trap that is stable with the right choice of parameters. The conditions for both types of traps can even be satisfied simultaneously.

Even in cases where maladaptive dynamics cannot be stabilized by increasing the net source-sink flow alone, they can be stabilized by an additional dispersal turnover $\alpha_C^+$ (case VI) if it induces a negative intrapatch feedback in both patch types, which is the case if $0<\hat{\omega}^{-+}<\omega^{+-}$ and $\hat{\omega}^{+-}<1<\omega^{-+}$. A perceptual and ecological trap can be simultaneously stable if $\phi^+>\mu^+$ and $\phi^->\mu^-$ and all other conditions for $\eta^k$ and $R_S^k$ are satisfied.

\section{\label{sec NA}Numerical results and larger food webs}

\begin{figure*}
\begin{subfigure}
    \centering
    \includegraphics[width=0.37\textwidth]{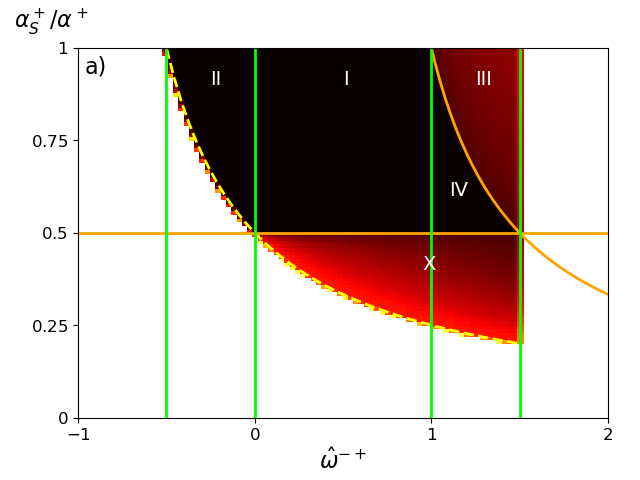}
\end{subfigure}
\begin{subfigure}
    \centering
    \includegraphics[width=0.37\textwidth]{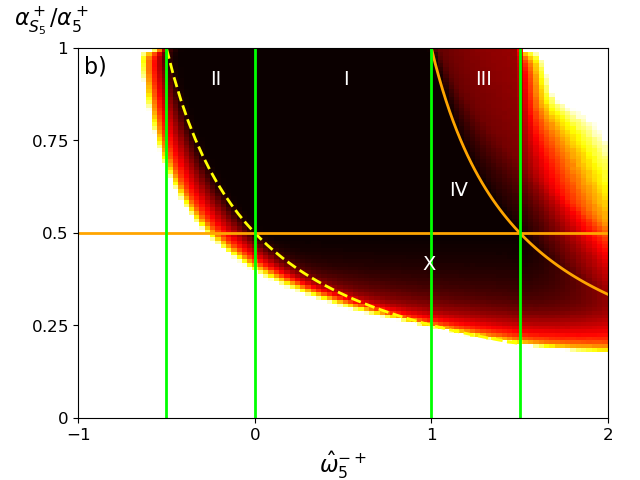}
\end{subfigure}
\begin{subfigure}
    \centering
    \includegraphics[width=0.37\textwidth]{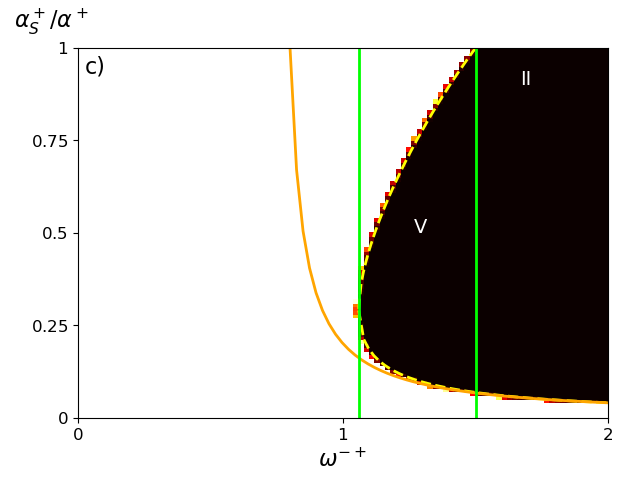}
\end{subfigure}
\begin{subfigure}
    \centering
    \includegraphics[width=0.37\textwidth]{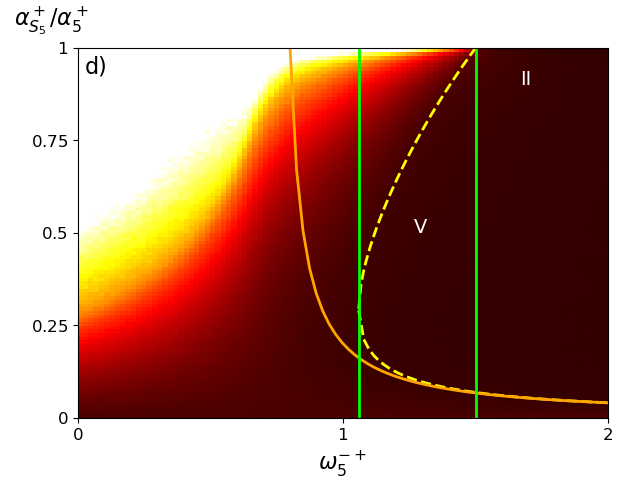}
\end{subfigure}
\begin{subfigure}
    \centering
    \includegraphics[width=0.37\textwidth]{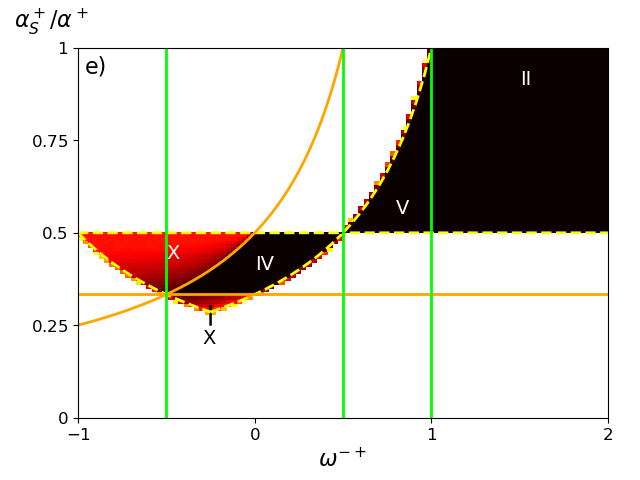}
\end{subfigure}
\begin{subfigure}
    \centering
    \includegraphics[width=0.37\textwidth]{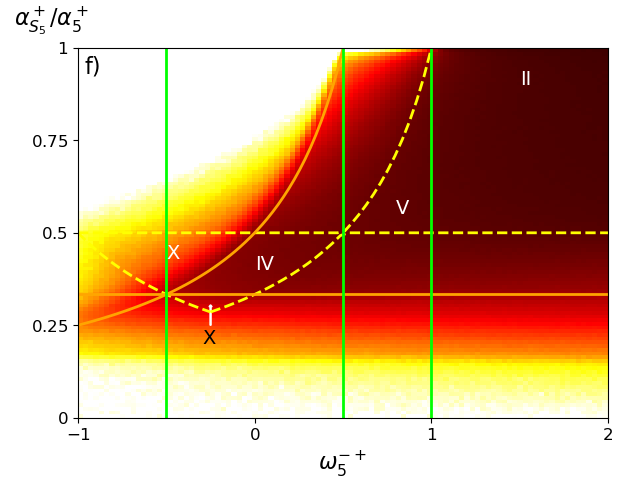}
\end{subfigure}
\begin{subfigure}
    \centering
    \includegraphics[width=0.37\textwidth]{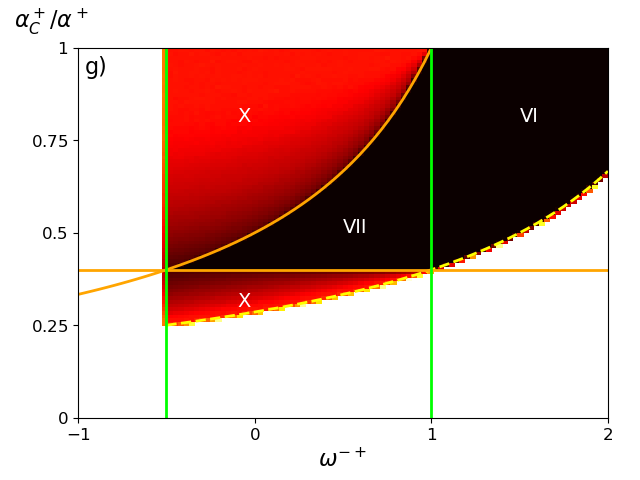}
\end{subfigure}
\begin{subfigure}
    \centering
    \includegraphics[width=0.37\textwidth]{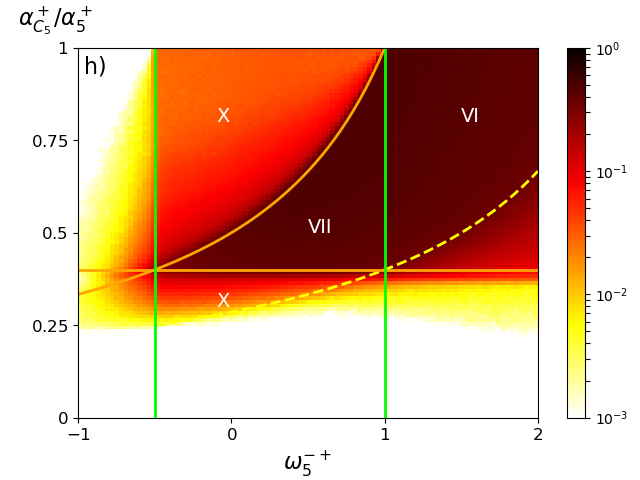}
\end{subfigure}
 \caption{The percentage of stable webs for metapopulations ((a), (c), (e), and (g)) or metacommunities with 10 species ((b), (d), (f), and (h)) in dependence of the normalized net source-sink flow $\alpha_{S_i}^+/\alpha_i^+$ or dispersal turnover $\alpha_{C_i}^+/\alpha_i^+$ and exponents of dispersal $\omega_i^{-+}$ or $\hat{\omega}_i^{-+}$. The different areas are marked with the corresponding stability criterion I-VII. Their boundaries (dashed yellow lines), the resistances (solid horizontal or curved orange lines), and transitions between cases I-VII that depend on $\omega_i^{+-}$ or $\hat{\omega}_i^{+-}$ (solid vertical green lines) are marked. (Color online)\label{plot results}}
\end{figure*}

To confirm the analytical results and to check how these translate to metacommunities with more species, we performed numerical analysis by varying the net source-sink flow $\alpha_{S_i}^+$ or the dispersal turnover $\alpha_{C_i}^+$. The parameter regions where the stability criteria I to VII are satisfied are indicated in Fig. \ref{plot results}. We evaluated the proportion of stable webs in dependence of the normalized turnover rates $\alpha_{S_i}^+/\alpha_i^k$ (with $\alpha_{C_i}^+=0$) or $\alpha_{C_i}^+/\alpha_i^k$ (with $\alpha_{S_i}^+=0$), with
\begin{equation}
 \alpha_i^k = \alpha_{P_i}^k + \alpha_{S_i}^k + \alpha_{C_i}^k\, .
\end{equation}
The second parameter that we varied is either $\omega_i^{+-}$ or $\hat{\omega}_i^{+-}$, as these two parameters play a central role in the stability conditions. For the metacommunities with several species these parameters were varied for only one species. The yellow dashed lines mark transitions where all eigenvalues become negative (see Appendices \ref{Appendix_NSSF} and \ref{Appendix_DT} for the formulas), and hence these are the boundaries of the stable regions. The solid horizontal and curved orange lines give the values of the resistances, and the vertical green lines indicate the transition between different stable regimes when changing $\omega_i^{+-}$ or $\hat{\omega}_i^{+-}$. These lines are marked in the plots for larger metacommunities ($N=10$) as well, such that we can compare the numerical results to the analytical results for one species. For more details of the numerical procedure (parameter values, methods used, etc.), see Appendix \ref{Appendix_NA}.

The black areas in Figs. \ref{plot results}(a), \ref{plot results}(c), \ref{plot results}(e), and \ref{plot results}(g) indicate parameter ranges where all systems are stable. These regions are delimited by the lines that were calculated analytically in the previous section. Region III in Fig. \ref{plot results}(a), however, has only $\approx 50\%$ stable systems. The reason is that the number of source and sink patches was chosen randomly, but only systems with more sources than sinks ($M^+>M^-$) are stable since then the additional eigenvalue \eqref{lambda add} is $\lambda_0^+$ and is negative in region III for the chosen parameter set. Otherwise, the additional eigenvalue is $\lambda_0^-$ which is positive since $\hat{\omega}^{-+}>\mu^-$.

Figs. \ref{plot results}(a), \ref{plot results}(e), and \ref{plot results}(g) include parameter regions marked with an X, where part of the systems are stable. These are regions where the topology of the source-sink network affects stability since stability depends on the smallest value of $\beta^2$, which we labeled $\beta_0^2$ in the previous section, and which we set to 0 in the analytical calculations to derive sufficient stability criteria (see Appendix \ref{Appendix_NSSF}). With this simplification, the affected regions of stability are delimited by the resistances, marked by the orange lines in the figure. This explains why the regions marked by an X occur when an orange line is crossed from a black region. Since $\beta_0^2$ must be larger further beyond the orange line for a system to be stable, the proportion of stable systems decreases with distance from this line.

When metacommunities of more than one species are evaluated (Figs. \ref{plot results}(b), \ref{plot results}(d), \ref{plot results}(f), and \ref{plot results}(h)) the stability criteria obtained for metapopulations with one species still play an important role, even though the black regions now become less dark, indicating that no longer 100\% of systems are stable.  On the other hand, parameter regions that were always unstable for the metapopulations, now include stable systems. This extension of parameter regions that allow for stable systems is most striking in \ref{plot results} (b).
Interestingly, in Fig. \ref{plot results}(f) and Fig. \ref{plot results}(h) orange resistance lines that were lying in the fully unstable regions in the single species case, now mark transitions between a higher and lower percentage of stable systems. In these regions stabilization was not possible for a single species due to positive interpatch feedbacks which cause a positive feedback loop. This suggests that the presence of other species might suppress positive feedback loops between source and sink populations.

\section{\label{discussion}Discussion and Conclusion}

In this paper, we have generalized the master stability function approach for metanetworks to systems with two types of patches. Combined with the generalized modeling approach, which allows to explore efficiently the linear stability of large networks, we have thus a powerful tool to evaluate the conditions under which large inhomogeneous metanetworks are stable. However, the advantages of the method come at the cost of limiting the model to steady states that are identical on all patches of the same type and to biregular patch topologies. Nevertheless, previous work shows that results obtained with such restrictions are usually more generally valid \cite{brechtel2019far}. 

We demonstrated the usefulness of the method by studying an explicit system, which is a metafoodweb on a spatial network of sources and sinks. Often, the dynamics of source-sink systems are either modelled only with one patch of each type \cite{holt1985,amarasekare2004,wu2020,matsumoto1995} or with explicit population dynamics \cite{holt1985,pulliam1988,amarasekare2004,arditi2015,wu2020,bansaye2013,matsumoto1995} or both. The choice of population dynamics equations fixes the values of most exponent parameters, for instance logistic growth in sources implies $\phi^+=1$, $\mu^+=2$. Often the sinks are modelled with a positive local feedback such that they are unable to maintain a population without immigration. The exponent of dispersal is also fixed in these models. This means that no variation of the exponents is taken into account, and therefore the insights about the stability of steady states in these studies is very limited. The study by P. Amarasekare \cite{amarasekare2004} varied the exponent $s$ of emigration rates from the source and found that if $s>1$ dispersal is stabilizing if large enough.
The same was done for logistic growth in sinks with an identical result. In the light of our findings, these results are merely a special case: Logistic growth in sources corresponds to $\phi^+=1$, and $s$ is the same exponent as our exponent $\omega^{-+}$, such that the condition $\omega^{-+}>\phi^+$ is satisfied. The other exponents are chosen as $\hat{\omega}_i^{-+}=0$ and $\mu^-=1$ and fulfill our condition $\hat{\omega}_i^{-+}<\mu^-$. 
 
Our approach thus allows a far broader investigation, not just for broad ranges of exponent parameters, but also for source-sink systems with more than two patches. Indeed, one source may be sustaining more than one sink and one sink can have more than one source as origin of its immigrating biomass \cite{tittler2006}. Hence a biregular set of patches poses a generalization of the often used two-patch approximation. In fact, there are natural topologies which are close to bipartite sets of patches. In addition to the home ranges and roaming regions of higher animals mentioned in the Introduction, further examples are given by ponds in deserts or dry lands, mainland-island structures, and landscapes sprinkled with lakes. Other examples with more than one patch of both types are mountains and valleys or lakes and rivers. More generally, each metafoodweb where per-capita growth rates are distributed heterogeneously can be seen as a network of sources and sinks because some patches are net exporters of biomass, while others are net importers.

We found that two distinct types of dispersal related turnover rates emerge naturally from the generalized model framework. One is the dispersal turnover $\alpha_{C_i}^+$ which corresponds to the amount of biomass which is exchanged between sources and sinks. The other is the net source-sink flow $\alpha_S^+$ which denotes the flow of biomass from sources to sinks and therefore the strength of source-sink dynamics.
We performed an analytical evaluation of a metacommunity with only one species as well as a numerical evaluation of metafoodwebs with many species in order to explore the capacity of both types of turnover rates to stabilize a system. In both situations, we found similar but not identical general conditions for such stabilizing effects. Stabilization can occur when rates become sufficiently large, but we also identified conditions under which a system is unstable for large and small rates but stable for an intermediate range of values of the rates. Such stabilizing effects of dispersal cannot occur in systems that are completely homogeneous \cite{brechtel2018}. 

We found two basic mechanisms which are stabilizing in source-sink dynamics for a single metapopulation, namely self-regulation (i.e., a decrease of growth rate with population size), and negative feedback loops between source and sink populations. 
The net source-sink flow $\alpha_S^+$ affects self-regulation through local dynamics as well as through dispersal, and its increase stabilizes the system under suitable conditions. But even if an increased net source-sink flow reduces the ability of a population to self-regulate, it can still have a stabilizing effect if it induces a negative feedback loop between source and sink populations. The stabilizing effect of these negative feedback loops depends on the spatial structure of the web, which is captured in the spectrum of singular values of the biregular network of patches.
Conversely, a positive feedback loop between source and sink populations of a single species can destabilize dynamics even if the net source-sink flow increases self-regulation. Our numerical analysis suggests that these positive intra-specific feedback loops may be suppressed by the presence of other species in source-sink dynamics. Though further research is needed for clear insights on that topic.

One particularly interesting finding is that dispersal can be stabilizing even if both patch types have a positive local feedback \eqref{local feedback}. This means that source-sink dynamics can stabilize metapopulations that are subject to strong local facilitation, which is equivalent to a positive local feedback. In particular, source-sink dynamics can be relevant for stabilizing not just sinks, but also sources. Sinks can provide mortality to source populations which would otherwise suffer from detrimental effects due to overcrowding. In contrast strong dispersal from sources to sinks may destabilize source populations (and thus possibly the whole source-sink system) due to the extent of losses \cite{amarasekare2004}. Underestimating the importance of sinks coupled to these sources might have catastrophic consequences for ecosystems. Hence it is important to evaluate the interaction between sources and sinks. Since exponent parameters are relatively easy to obtain from data \cite{fell1985,yeakel2011} the findings of our study can be of practical use for identifying appropriate measures that preserve the stability of source-sink systems. 
 
Another striking finding is that there exist generic conditions under which ecological and perceptual traps can be stable, even though dispersal is maladaptive in these situations. So far traps have mostly been seen as detrimental \cite{hale2016}, but may provide a mechanism to stabilize metacommunities by limiting strong Allee effects. Clearly further research is needed to identify the effects of traps on metacommunity persistence. 

The listed general results are only a small part of what can still be achieved with the method. We expect that the formalism can be used to evaluate properties of other bipartite metanetworks, such as mutualistic ecological networks.

\section*{Acknowledgements}
T.G. was supported by the Ministry for Science and Culture of Lower Saxony and the Volkswagen Foundation through the “Nieders\"{a}chsisches Vorab” grant program (Grant No. ZN3285)
This project was funded by the Deutsche Forschungsgemeinschaft (DFG, German Research Foundation) - Grant No. DR300/15.

\appendix
\section{Calculation of the generalized model of source-sink dynamics}
\label{Appendix_GM}

We start with Eq. \eqref{general model},
\begin{eqnarray}
    \dot{X}_i^k = &&G_i^k(\bm{X}^k) - M_i^k(\bm{X}^k) \nonumber\\
                  &&+ \sum_{l=1}^M E_i^{kl}(\bm{X}^k,\bm{X}^l) - \sum_{l=1}^M E_i^{lk}(\bm{X}^l,\bm{X}^k)
\end{eqnarray}
and rewrite it as a generalized model following the framework in \cite{gross2006}. We assume that there exists a steady state $(\bm{X}^k)^*$. All population densities and functions are normalized to their respective values at the steady state, and hence we define the normalized population densities $\bm{x}^k$ with $x_i^k=X_i^k/(X_i^k)^*$ and the normalized functions $g_i^k(\bm{x}^k)=G_i^k(\bm{X}^k)/(G_i^k)^*$, $m_i^k(\bm{x}^k)=M_i^k(\bm{X}^k)/(M_i^k)^*$, and $e_i^{kl}(\bm{x}^k,\bm{x}^l)=E_i^{kl}(\bm{X}^k,\bm{X}^l)/(E_i^{kl})^*$. Inserting these into the model equation we find 
\begin{eqnarray}
\dot{x}_i^k =&&  \frac{(G_i^k)^*}{(X_i^k)^*} g_i^k(\bm{x}^k) - \frac{(M_i^k)^*}{(X_i^k)^*} m_i^k(\bm{x}^k) \\
		 &&+ \sum_{l=1}^M  \frac{(E_i^{kl})^*}{(X_i^k)^*} e_i^{kl}(\bm{x}^k, \bm{x}^l) 
		 - \sum_{l=1}^M   \frac{(E_i^{lk})^*}{(X_i^k)^*} e_i^{lk}(\bm{x}^l, \bm{x}^k)\; . \nonumber
\end{eqnarray}
At the steady state we have $\dot{x}_i^k=0$ and $x_i^k=1$, $g_i^k(\bm{x}^k)=1$, $m_i^k(\bm{x}^k)=1$, $e_i^{lk}(\bm{x}^l, \bm{x}^k)=1$. Hence the sum of the coefficients of all gain terms has to be equal to the sum of the coefficients of all loss terms. This allows us to define the total biomass turnover rate $\alpha_i^k$ as 
\begin{equation}
\alpha_i^k = \frac{(G_i^k)^*}{(X_i^k)^*} + \sum_{l=1}^M  \frac{(E_i^{kl})^*}{(X_i^k)^*}
= \frac{(M_i^k)^*}{(X_i^k)^*} + \sum_{l=1}^M   \frac{(E_i^{lk})^*}{(X_i^k)^*}\; .
\end{equation}
Further we define scale parameters for the gain
\begin{eqnarray}
\nu_i^k &=& \frac{1}{\alpha_i^k}\frac{(G_i^{k})^*}{(X_i^{k})^*}\; ,\nonumber \\
\tilde{\nu}_i^{k} 	&=& 1 - \nu_i^k = \sum_{l=1}^M \tilde{\nu}_i^{kl} \tilde{\nu}_i^k = \frac{1}{\alpha_i^k} \sum_{l=1}^M \frac{(E_i^{kl})^*}{(X_i^{k})^*}\; , 
\end{eqnarray}
and for the loss terms
\begin{eqnarray}
\rho_i^k &=& \frac{1}{\alpha_i^k} \frac{(M_i^{k})^*}{(X_i^{k})^*}\; ,\nonumber\\
\tilde{\rho}_i^k &=& 1 - \rho_i^k = \sum_{l=1}^M \tilde{\rho}_i^{lk} \tilde{\rho}_i^k = \frac{1}{\alpha_i^k} \sum_{l=1}^M \frac{(E_i^{lk})^*}{(X_i^{k})^*}\; .
\end{eqnarray}
These denote the contribution of the respective gain (loss) term to the total gain (loss) of species $i$ on patch $k$. There are $d^k$ nonzero parameters $\tilde{\nu}_i^{kl}$ and $\tilde{\rho}_i^{lk}$, respectively, and we set them to equal values to satisfy the constraints imposed in Sec. \ref{sec MSF}. These parameters denote the contribution of each link to the total gain ($\tilde{\nu}_i^{kl}$) or loss ($\tilde{\rho}_i^{lk}$) in patch $k$. Hence we set $\tilde{\nu}_i^{kl}=\tilde{\rho}_i^{lk}=1/d^k$. We thus find 
\begin{eqnarray}
\dot{x}_i^k =&& \alpha_i^k \bigg[\nu_i^k g_i^k(\bm{x}^k) -\rho_i^k m_i^k(\bm{x}^k) \label{normalized model A}\\
		 &&+ \frac{1-\nu_i^k}{d^k} \sum_{l=1}^M  e_i^{kl}(\bm{x}^k, \bm{x}^l) - \frac{1-\rho_i^k}{d^k} \sum_{l=1}^M   e_i^{lk}(\bm{x}^l, \bm{x}^k)\bigg]\; . \nonumber
\end{eqnarray}
We will choose a form of Eq. \eqref{normalized model A} that naturally reflects the differences between sources and sinks. Local production is larger than local mortality in sources and vice versa in sinks. Similarly, the rate of emigration is larger than that of immigration for sources and smaller than that of immigration for sinks. Then the difference between growth and mortality is equal to the difference between immigration and emigration in both patch types. By making appropriate parameter substitutions, taking into account that all sources have the same set of parameters, and all sinks have the same set of parameters, we can write Eq. \eqref{normalized model A} as
\begin{eqnarray}
\dot{x}_i^{k_+} =& \alpha_{P_i}^{+} &\bigg[g_i^{k_+}(\bm{x}^{k_+})  - m_i^{k_+}(\bm{x}^{k_+}) \bigg] \nonumber\\
	    &+ \alpha_{S_i}^{+} &\bigg[g_i^{k_+}(\bm{x}^{k_+}) - \frac 1 {d^+}\sum_{k_-} e_i^{{k_-}{k_+}}(\bm{x}^{k_-}, \bm{x}^{k_+})\bigg] \nonumber\\
	    &+ \alpha_{C_i}^{+} &\bigg[\frac 1 {d^+}\sum_{k_-}  e_i^{{k_+}{k_-}}(\bm{x}^{k_+}, \bm{x}^{k_-}) \nonumber\\
	    &&- \frac 1 {d^+}\sum_{k_-}  e_i^{{k_-}{k_+}}(\bm{x}^{k_-}, \bm{x}^{k_+})\bigg] \label{normalized source A}
\end{eqnarray}
for all sources and 
\begin{eqnarray}
\dot{x}_i^{k_-} =& \alpha_{P_i}^{-} &\bigg[g_i^{k_-}(\bm{x}^{k_-})  - m_i^{k_-}(\bm{x}^{k_-}) \bigg] \nonumber\\
	    &+ \alpha_{S_i}^{-} &\bigg[\frac 1{d^-}\sum_{k_+}  e_i^{{k_-}{k_+}}(\bm{x}^{k_-}, \bm{x}^{k_+}) - m_i^{k_-}(\bm{x}^{k_-}) \bigg] \nonumber\\
	    &+ \alpha_{C_i}^{-} &\bigg[\frac 1{d^-}\sum_{k_+}  e_i^{{k_-}{k_+}}(\bm{x}^{k_-}, \bm{x}^{k_+}) \nonumber\\
	    &&- \frac {1}{d^-}\sum_{k_+} e_i^{{k_+}{k_-}}(\bm{x}^{k_+}, \bm{x}^{k_-})\bigg] \label{normalized sink A}
\end{eqnarray}
for all sinks.

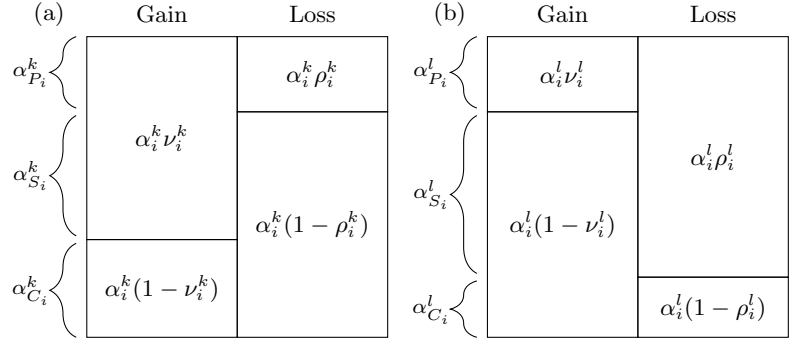
\begin{figure*}
\begin{subfigure}
\centering
    \begin{tikzpicture}
    \draw (-2.5,2.3) node{(a)};
    \draw (-2,-2) rectangle (0,-0.7);
    \draw (-2,-0.7) rectangle (0,2);
    \draw (0,-2) rectangle (2,1.);
    \draw (0,1.) rectangle (2,2);
    \draw (-1,0.65) node{$\alpha_i^k \nu_i^k$};
    \draw (-1,-1.35) node{$\alpha_i^k (1-\nu_i^k)$};
    \draw (1,1.5) node{$\alpha_i^k \rho_i^k$};
    \draw (1,-0.5) node{$\alpha_i^k (1-\rho_i^k)$};
    \draw (-1,2.3) node{Gain};
    \draw (1,2.3) node{Loss};
    \draw [decorate,decoration={brace,amplitude=10pt},xshift=-4pt,yshift=0pt] (-2,-2) -- (-2,-0.75) node [black,midway,xshift=-0.6cm] {\footnotesize $\alpha_{C_i}^k$};
    \draw [decorate,decoration={brace,amplitude=10pt},xshift=-4pt,yshift=0pt] (-2,-0.65) -- (-2,.95) node [black,midway,xshift=-0.6cm] {\footnotesize $\alpha_{S_i}^k$};
    \draw [decorate,decoration={brace,amplitude=10pt},xshift=-4pt,yshift=0pt] (-2,1.05) -- (-2,2) node [black,midway,xshift=-0.6cm] {\footnotesize $\alpha_{P_i}^k$};
    \end{tikzpicture}
\end{subfigure}
\begin{subfigure}
\centering
    \begin{tikzpicture}
    \draw (-2.5,2.3) node{(b)};
    \draw (-2,-2) rectangle (0,1);
    \draw (-2,1) rectangle (0,2);
    \draw (0,-2) rectangle (2,-1.2);
    \draw (0,-1.2) rectangle (2,2);
    \draw (-1,1.5) node{$\alpha_i^l \nu_i^l$};
    \draw (-1,-0.5) node{$\alpha_i^l (1-\nu_i^l)$};
    \draw (1,0.4) node{$\alpha_i^l \rho_i^l$};
    \draw (1,-1.6) node{$\alpha_i^l (1-\rho_i^l)$};
    \draw (-1,2.3) node{Gain};
    \draw (1,2.3) node{Loss};
    \draw [decorate,decoration={brace,amplitude=10pt},xshift=-4pt,yshift=0pt] (-2,-2) -- (-2,-1.25) node [black,midway,xshift=-0.6cm] {\footnotesize $\alpha_{C_i}^l$};
    \draw [decorate,decoration={brace,amplitude=10pt},xshift=-4pt,yshift=0pt] (-2,-1.15) -- (-2,.95) node [black,midway,xshift=-0.6cm] {\footnotesize $\alpha_{S_i}^l$};
    \draw [decorate,decoration={brace,amplitude=10pt},xshift=-4pt,yshift=0pt] (-2,1.05) -- (-2,2) node [black,midway,xshift=-0.6cm] {\footnotesize $\alpha_{P_i}^l$};
    \end{tikzpicture}
 \end{subfigure}
\caption{Illustration of the relation between the original scale parameters in Eqs. \eqref{normalized model A} and the turnover rates introduced in Eqs. \eqref{normalized source} and \eqref{normalized sink} for (a) sources and (b) sinks. \label{alpha bars}}
\end{figure*}

Fig. \ref{alpha bars} illustrates the relationship between the scale parameters and the turnover rates $\alpha_{P_i}^k$, $\alpha_{S_i}^k$, and $\alpha_{C_i}^k$.
In addition we take into account that the total biomass outflow  from patch $l$ into patch $k$ has to be identical to the biomass inflow into patch $k$ from patch $l$, requiring that
\begin{equation}
\label{consistency}
  \alpha_i^l (1-\rho_i^l) \frac{X_i^{l*}}{d^l} = \alpha_i^k (1-\nu_i^k) \frac{X_i^{k*}}{d^k}\; .
\end{equation}
Applying condition \eqref{consistency} to the situation where $l$ is the sink and $k$ is the source, and vice versa, we obtain 
\begin{equation}
  q \alpha_{C_i}^+ = \alpha_{C_i}^- \quad \hbox{ and } \quad q\alpha_{S_i}^+ = \alpha_{S_i}^- 
\end{equation}
with 
\begin{equation} \label{q}
 q = \frac{X_i^{+*}}{X_i^{-*}} \frac{d^-}{d^+}\; .
\end{equation}
By using these relations we arrive at the generalized model given in Eqs.~\eqref{normalized source} and \eqref{normalized sink}. The meaning and realistic range of values of all generalized parameters are given in TABLE \ref{table generalized parameters}.

\begin{table*}
\begin{ruledtabular}
\begin{tabular}{lll}
Parameter & Interpretation & realistic range\\
\hline
Turnover 	& &\\
\hline
$\alpha_i^k$        &Total biomass turnover& $> 0$ \\
$\alpha_{P_i}^k$    &Purely local biomass turnover consisting of parts of local growth and mortality& $> 0$\\
$\alpha_{S_i}^k$    &Biomass turnover due to source sink dynamics& $\geq 0$\\
$\alpha_{C_i}^k$    &Biomass turnover due to pure dispersal dynamics& $\geq 0$\\
\hline
Scale 		& &\\
\hline
$\nu_i^k$	&Fraction of local growth $G_i^k(\bm{x}^k)$ to total gain.&[0,1]\\
$\rho_i^k$	&Fraction of local mortality $M_i^k(\bm{x}^k)$ to total loss.&[0,1]\\
$1-\nu_i^k$			&Fraction of immigration $\sum_l E_i^{kl}(\bm{x}^k, \bm{x}^l)$ to total gain.&[0,1]\\
$1-\rho_i^k$			&Fraction of emigration $\sum_l E_i^{lk}(\bm{x}^l, \bm{x}^k)$ to total loss.&[0,1]\\
$\nu_i^{kl}$		&Fraction of immigration $E_i^{kl}(\bm{x}^k, \bm{x}^l)$ along the link from $l$ to $k$&[0,1]\\
$\rho_i^{kl}$		&Fraction of emigration $E_i^{lk}(\bm{x}^l, \bm{x}^k)$ along the link from $l$ to $k$&[0,1]\\
\hline
Exponent	& &\\
\hline	
$\phi_i^k$			   &Exponent of $G_i^k(\bm{x}^k)$ with respect to population $i$&[0,2]\\
$\phi_{ij}^k$		   &Exponent of $G_i^k(\bm{x}^k)$ with respect to population $j$&[0,2]\\
$\mu_i^k$			   &Exponent of $M_i^k(\bm{x}^k)$ with respect to population $i$&[0,2]\\
$\mu_{ij}^k$		   &Exponent of $M_i^k(\bm{x}^k)$ with respect to population $j$&[0,2]\\
$\hat{\omega}_i^{kl}$  &Exponent of $E_i^{kl}(\bm{x}^k, \bm{x}^l)$ with respect to population $i$ in target patch $k$&[-2,2]\\
$\omega_i^{kl}$		   &Exponent of $E_i^{kl}(\bm{x}^k, \bm{x}^l)$ with respect to population $i$ in starting patch $l$&[-2,2]\\	
$\hat{\kappa}_{ij}^{kl}$ &Exponent $E_i^{kl}(\bm{x}^k, \bm{x}^l)$  with respect to population $j$ in target patch $k$&[-2,2]\\
$\kappa_{ij}^{kl}$		&Exponent of $E_i^{kl}(\bm{x}^k, \bm{x}^l)$ with respect to population $j$ in starting patch $l$ &[-2,2]\\
\end{tabular}
\end{ruledtabular}
\caption{Generalized parameters of species $i$ and their ecological interpretation. The exponent parameters are elasticities of the respective function with respect to population sizes. $k=+,-$ is associated with either sources (+) or sinks (-).\label{table generalized parameters}}
\end{table*}

\section{Analytical calculation of the eigenvalues of the Jacobian for one species}
\label{Appendix_EV}

In the following we calculate the eigenvalues of the reduced Jacobian $\bm{j}$ \eqref{reduced jacobian} for one species. Since this is a $2\times 2$ matrix, its eigenvalues $\lambda_i$ are given by its trace $\text{tr}(\bm{j})$ and determinant $\text{det}(\bm{j})$ by
\begin{equation}
 \lambda_i = \frac{\text{tr}(\bm{j})}{2} \pm \sqrt{\frac{\text{tr}(\bm{j})^2}{4} - \text{det}(\bm{j})}\; .
\end{equation}
In our system, the trace of $\bm{j}$ is
\begin{eqnarray}
 \text{tr}(\bm{j}) &=& P^+ + P^- - d^+ C^+ - d^- C^- \nonumber \\
                    &=& \alpha_P^+ (\phi^+ - \mu^+) +\alpha_P^- (\phi^- - \mu^-) \nonumber \\
                    &&+ \alpha_S^+ (\phi^+ - \omega^{-+} +q\hat{\omega}^{-+} - q\mu^-) \nonumber \\
                    &&+ \alpha_C^+ (\hat{\omega}^{+-} - \omega^{-+} + q\hat{\omega}^{-+} - q\omega^{+-})\, , \label{trace}
\end{eqnarray}
and the determinant is
\begin{equation}
 \text{det}(\bm{j}) = (P^+- d^+C^+)(P^- - d^-C^-)- \beta^2 \hat{C}^+ \hat{C}^- \; .\label{det}
\end{equation}
Stability is given if $\text{tr}(\bm{j})<0$ and $\text{det}(\bm{j})>0$. In the following we will use these conditions to determine stability of a given system.

\section{Influence of the net source-sink flow on stability}
\label{Appendix_NSSF}
\subsection{Stability for large net source-sink flows}
\label{Appendix_LNSSF}

If making $\alpha_S^+$ arbitrarily large shall have a stabilizing effect, then the trace of the reduced Jacobian $\bm{j}$ must become negative and the determinant positive for sufficiently large $\alpha_S^+$. According to Eqs. \eqref{trace} and \eqref{det}, the trace is linear in $\alpha_S^+$ and the determinant is a quadratic function in $\alpha_S^+$. Further, the additional eigenvalue $\lambda_0^k$ in \eqref{lambda add} must be negative for sufficiently large $\alpha_S^+$. Hence, the one-species source-sink dynamics are linearly stable for large $\alpha_S^+$ if either
\begin{equation}
\label{condition 0 eigenvalues +}
 \omega^{-+}>\phi^+ \quad \text{if}\quad M^+>M^- 
\end{equation}
 or
\begin{equation}
\label{condition 0 eigenvalues -}
 \mu^->\hat{\omega}^{-+} \quad \text{if} \quad  M^->M^+\, ,
\end{equation}
and
\begin{equation}
\label{condition trace s1}
 \frac{\partial \text{tr}(\bm{J})}{\partial \alpha_S^+} = \phi^+ - \omega^{-+} +q(\hat{\omega}^{-+} - \mu^-) < 0
\end{equation}
and
\begin{eqnarray}
 \frac{\partial^2 \text{det}(\bm{J})}{\partial (\alpha_S^+)^2} = &q [(\phi^+-\omega^{-+}) (\hat{\omega}^{-+}-\mu^-) \nonumber\\
 &+ \frac{\beta^2}{d^+d^-} \hat{\omega}^{-+} \omega^{-+}] > 0\; . \label{condition determinant s1}
\end{eqnarray}

The topology of the patch network affects stability only through the singular values $\beta$. Condition \eqref{condition determinant s1} has to be satisfied for all $\beta \in S$. For the special case that $M^+=M^-$ there is no additional eigenvalue $\lambda_0^k$ and thus no condition \eqref{condition 0 eigenvalues +} or \eqref{condition 0 eigenvalues -}. 

If $\omega^{-+}<0$, then condition \eqref{condition 0 eigenvalues +} cannot be satisfied. Further, conditions \eqref{condition 0 eigenvalues -}, \eqref{condition trace s1}, and \eqref{condition determinant s1} cannot be satisfied simultaneously, and hence we find: 
\begin{quote}
\textit{A large net source-sink flow $\alpha_S^+$ cannot stabilize the system if emigration and density in sources are negatively correlated, i.e., if $\omega^{-+}<0$.}
    \end{quote}

We therefore assume in the remainder of this section that $\omega^{-+}\geq 0$, and we identify several sets of conditions under which relations \eqref{condition 0 eigenvalues +} to \eqref{condition determinant s1} can be satisfied for all $\beta \in S$. 

We obtain different stabilizing conditions depending on whether \eqref{condition 0 eigenvalues +} and \eqref{condition 0 eigenvalues -}  are both satisfied simultaneously or only one of them, i.e., if the net source-sink flow induces a negative intrapatch feedback in either both or only one patch type. In the first case, either a negative ($-\hat{\omega}^{-+}\hat{\omega}^{-+}<0$) or a positive interpatch feedback ($\hat{\omega}^{-+}\hat{\omega}^{-+}>0$) is possible; in the second case a negative interpatch feedback is needed (-$\hat{\omega}^{-+}\hat{\omega}^{-+}<0$). Hence we find three ways $\alpha_S^+$ can have a stabilizing effect, which result in the cases I-III in Sec. \ref{sec NSSF}.

\subsubsection{Case I: the net source-sink flow induces negative intrapatch feedbacks and a negative interpatch feedback}

If conditions  \eqref{condition 0 eigenvalues +} and \eqref{condition 0 eigenvalues -} are satisfied and additionally $\hat\omega^{-+}> 0$, the two intrapatch feedbacks are negative as well as the interpatch feedback (see Sec. \ref{sec NSSF}).

$\hat\omega^{-+}> 0$ means that condition \eqref{condition determinant s1} is satisfied for all $\beta$, and we need no further restrictions on the parameters. Thus the negative intrapatch feedbacks \eqref{intrapatch feedback S} induced by $\alpha_S$ are sufficient for stabilization. In particular condition \eqref{condition 0 eigenvalues -} assures that the positive effect of immigration on sink population growth is limited by the density-dependent increase in mortality for population sizes above the steady state. A large enough net source-sink flow $\alpha_S^+$ can thus potentially stabilize a sink that would be unstable otherwise. This is the well-known rescue effect \cite{brown1977,eriksson2014}. 

\subsubsection{Case II: the net source-sink flow induces negative intrapatch feedbacks and a positive interpatch feedback}

Again conditions  \eqref{condition 0 eigenvalues +} and \eqref{condition 0 eigenvalues -} are satisfied, but now $\hat\omega^{-+}< 0$. This means that condition \eqref{condition determinant s1} has a negative last term, the absolute value of which is largest for  $\beta^2 = d^+d^-$. Condition \eqref{condition determinant s1} can be rewritten in this case as a condition for $|\hat{\omega}^{-+}|$,
\begin{equation}
\label{result alpha s2}
|\hat{\omega}^{-+}| < {\mu^-} \left(\frac{\omega^{-+}}{\phi^+}-1\right)\, 
 \end{equation}
which must be satisfied for stability. Otherwise a detrimental positive feedback loop is present.

\subsubsection{Case III: the net source-sink flow induces a positive intrapatch feedback in one patch type and a negative one in the other}
 
Now we consider the case that only one of the conditions  \eqref{condition 0 eigenvalues +} and \eqref{condition 0 eigenvalues -} is satisfied. In this case \eqref{condition determinant s1} can only be satisfied if $\hat{\omega}^{-+}>0$. If we denote with $\beta_0$ the smallest value of all $\beta \in S$, condition \eqref{condition determinant s1} now becomes
  \begin{equation}\label{condition M+-}
|(\phi^+-\omega^{-+})   (\hat\omega^{-+}-\mu^-)|   < \frac{\beta_0^2}{d^+d^-}   \hat\omega^{-+}\omega^{-+}\, .
  \end{equation}
Furthermore, condition \eqref{condition trace s1} becomes
 \begin{equation}
 \label{condition M+- 2}
 \frac{|\phi^+ -\omega^{-+}|}{q|\hat\omega^{-+}-\mu^-|} \gtrless 1 \quad \text{if}\quad M^- \lessgtr M^+\, .
 \end{equation}
In both cases the less numerous patch type has a positive diagonal entry in the Jacobian when the net source-sink flow $\alpha_S^+$ is large. Hence a positive intrapatch feedback is induced which has to be smaller than the negative intrapatch feedback which is induced in the more numerous patch (see \eqref{condition M+- 2}). As discussed before this is destabilizing due to a nonlocal demographic Allee effect. This destabilizing effect can be countered by a negative interpatch feedback ($-\omega^{-+}\hat{\omega}^{-+}<0$). The induced feedback loop has to be strong enough to suppress the demographic Allee effect, which is reflected by condition \eqref{condition M+-}.

\subsection{Stabilization for intermediate net source-sink flows}
\label{Appendix_INSSF}

In the following, we will consider the case that there is an unstable steady state at $\alpha_S^+=0$ and at $\alpha_S^+ \to \infty$, and we will demonstrate that there can be an intermediate interval of values of the net source-sink flow $\alpha_S^+$ for which the dynamics are stable. 
We will explicitly give three examples of parameter sets which satisfy the conditions for stability. Again we assume that $\alpha_C^+=0$ for the sake of simplicity. We write the determinant \eqref{det} as
\begin{equation}
\label{det new}
 \text{det}(\alpha_S^+) = A (\beta^2) (\alpha_S^+)^2 + B \alpha_S^+ + C
\end{equation}
and the trace \eqref{trace} as
\begin{equation} \label{trace new}
 \text{tr}(\alpha_S^+) = D \alpha_S^+ + E\, ,
\end{equation}
which means that 
\begin{eqnarray} \label{A(beta)}
    A (\beta^2) =& q[(\phi^+-\omega^{-+})(\hat{\omega}^{-+}-\mu^-) \nonumber\\
    &+ \frac{\beta^2}{d^+d^-}\omega^{-+}\hat{\omega}^{-+}]\\
    B =& \alpha_P^- (\phi^+-\omega^{-+})(\phi^- - \mu^-)\nonumber\\
    &+q\alpha_P^+ (\hat{\omega}^{-+}-\mu^-) (\phi^+-\mu^+)\label{B}\\
    C =& \alpha_P^+\alpha_P^-(\phi^+-\mu^+)(\phi^--\mu^-)\label{C}\\
    D =& \phi^+-\omega^{-+} + q(\hat{\omega}^{-+}-\mu^-) \label{D}\\
    E =& \alpha_P^+(\phi^+-\mu^+) + \alpha_P^-(\phi^--\mu^-) \label{E}\; .
\end{eqnarray}
If the steady state shall be unstable when dispersal is absent ($\alpha_S^+=0$), then at least one local feedback must be positive, and thus either $\phi^+-\mu^+<0$ or $\phi^- - \mu^-<0$ must be violated [see eq. \eqref{result local}]. Further, if the steady state shall be unstable for large  $\alpha_S^+$, then condition  \eqref{condition trace s1} ($D<0$) or condition \eqref{condition determinant s1} ($A(\beta^2)>0$) or both must be violated. Here we focus on the violation of \eqref{condition determinant s1} and assume that $A(\beta^2)<0$ for at least one $\beta \in S$.  (In fact, if $A(\beta^2)>0$ for all $\beta \in S$, then the conditions required to obtain an intermediate range of stability appear difficult to satisfy, as a numerical exploration of selected parameter combinations did not yield any positive results.)

Now, a straightforward way of obtaining a range of $\alpha_S^+$ values outside of which the system is unstable is when the determinant \eqref{det} has two positive roots, so that the determinant is positive within and negative outside these two roots. This means that $B>0$ and $C<0$ [see Fig. \ref{A minimal}(a)].  If an intermediate region shall be stable, then the determinant has to be positive for all $\beta \in S$ within this region. Since $\beta$ enters the determinant only via $A$, the determinant is positive for all $\beta \in S$ in the intermediate region if it is positive for the value $\beta^2$ for which $A(\beta^2)$ is minimal.

Which value of $\beta^2$ minimizes $A$, depends on the sign of $\hat{\omega}^{-+}\omega^{-+}$ [see \eqref{A(beta)}],
\begin{equation}
    \beta^2 = 
     \begin{cases}
       d^+d^- &\quad\text{if}\;  -\hat{\omega}^{-+}\omega^{-+}>0 \\
       \beta_0^2 &\quad\text{if}\;  -\hat{\omega}^{-+}\omega^{-+} < 0 \, .
     \end{cases} \label{betacases}
\end{equation}
We assume in the following for simplicity that the lowest singular value is $\beta_0=0$. The conditions for stability that we will find with this assumption are sufficient conditions. The parameter regions that show stability can therefore be somewhat larger than the ones calculated by us. 

Besides the conditions $B>0$, $C<0$, and $A<0$ for at least one $\beta$, we additionally require the condition
\begin{equation}
\label{condition roots}
  B^2-4AC > 0  
\end{equation}
in order to obtain two positive roots of the determinant \eqref{det new}. Let us write the expression \eqref{B} as $B=B_1+B_2$. In the case $\beta=0$ we have $B^2-4AC = (B_1-B_2)^2 >0$ such that we need no further constraints on $A,B$, and $C$. If $\beta > 0$, then the condition $B^2-4AC > 0 $ is still satisfied if $-\hat{\omega}^{-+}\omega^{-+}\le 0 $. But if $-\hat{\omega}^{-+}\omega^{-+}>0 $ [in which case $\beta^2=d^+d^-$ minimizes $A(\beta)$, see \eqref{betacases}], then the requirement that $B^2-4AC > 0$ must be met explicitly. We rewrite this condition (setting $\beta^2=d^+d^-$) as 
\begin{equation}\label{B1B2C}
    (B_1-B_2)^2 > 4 |C| |\hat{\omega}^{-+}\omega^{-+}|.
\end{equation}
A comparison with the explicit expressions for $B_1$ and $B_2$ in \eqref{B} shows that this inequality can be satisfied by increasing $\alpha_P^-$ or $\alpha_P^+$ or $q$ or $|\phi^+ - \mu^+|$ or $|\phi^- - \mu^-|$ sufficiently far while keeping the signs of the other terms $A$ to $E$ fixed. One of these possibilities will always be available without violating the conditions imposed on the signs of the terms or the other conditions for stability. This will be clear when looking at the examples below.

If the model shall be stable between the two roots of the determinant, then we have to satisfy two more conditions, namely that the trace \eqref{trace new} and the additional eigenvalue given by \eqref{lambda add} are negative. In order to specify constraints on the parameters that ensure these conditions, it is useful to define the source's and sink's resistances $R_S^+$ and $R_S^-$ to changes in the net-source sink flow $\alpha_S^+$ as
\begin{equation}
    R_S^+ = -\alpha_P^+  \frac{\phi^+ - \mu^+}{\phi^+-\omega^{-+}} \, , \quad R_S^- = -\frac{\alpha_P^-}{q}  \frac{\phi^- - \mu^-}{\hat{\omega}^{-+}-\mu^-} \, .
\end{equation}
They are the roots of the determinant \eqref{det new} for $\beta_0^2=0$ if the above-mentioned conditions ($A<0$, $B>0$, $C<0$) are satisfied. In two of our three examples below, we will have $-\hat{\omega}^{-+}\omega^{-+}<0$ such that $\beta^2=\beta^2=0$ minimizes $A$, and $R_S^+$ and $R_S^-$ are the two values of $\alpha_S^+$ that delimit the stable region.

The additional eigenvalue given by \eqref{lambda add} is
\begin{equation} \label{lambdacases}
    \lambda_0^k = 
     \begin{cases}
       \alpha_S^+ (\phi^+ - \omega^{-+}) + \alpha_P^+(\phi^+ - \mu^+) \; \text{if}\;  M^+>M^- \\
       \alpha_S^+ q(\hat{\omega}^{-+} - \mu^-) + \alpha_P^-(\phi^- - \mu^-) \; \text{if}\;  M^->M^+.
     \end{cases}
\end{equation}
Both expressions are linear in $\alpha_S^+$ and go through zero for $\alpha_S^+=  R_S^+$  if $M^+>M^-$ and for $\alpha_S^+= R_S^-$ if $M^+<M^-$. A larger resistance therefore means that the value of $\alpha_S^+$ required to change the sign of $\lambda_0^k$ is larger, i.e., the stability (or instability) of the uncoupled (i.e., $\alpha_S^+=0$) system is more resistant to an increase of $\alpha_S^+$. (And a negative value of the resistance means that an increase of $\alpha_S^+$ cannot change the sign of $\lambda_0^k$ at all since $\alpha_S^+$ cannot be negative.) 

The following three examples are chosen such that, by fixing the signs of the slope and intercepts of the additional eigenvalue, we can make sure that $\lambda_0^k$ is negative between the two roots of the determinant. Similarly, by fixing the signs of $D$ and $E$ in the trace \eqref{trace new} of the Jacobian, we can make sure that it is negative between the two roots of the determinant. For the first two examples the net source-sink flow $\alpha_S^+$ induces intrapatch feedbacks \eqref{intrapatch feedback S} with opposite signs. Those correspond to the case IV found in Sec. \ref{sec NSSF}. For the third example $\alpha_S^+$ induces a negative intrapatch feedbacks in both patch types, corresponding to case V. Then the destabilization for large $\alpha_S^+$ occurs due to a positive feedback loop ($-\hat{\omega}^{-+}\omega^{-+}>0$).

\subsubsection{Case IV: the net source-sink flow induces intrapatch feedbacks with opposite signs}

An example for how the conditions $A(\beta^2)<0$, $B>0$, $C<0$, and \eqref{condition roots} can be satisfied is given by  $\phi^+ - \mu^+<0$, $\phi^- - \mu^->0$, $\phi^+ - \omega^{-+}>0$, $q(\hat{\omega}^{-+} - \mu^-)<0$, $R_S^+>R_S^-$, and $-\hat{\omega}^{-+}\omega^{-+} \leq 0$.
According to \eqref{A(beta)} $A$ is minimal if $\beta^2=\beta_0^2=0$. If we compare the choice of parameters with the expressions \eqref{A(beta)} to \eqref{E}, then it is straightforward to see that $A(0)<0$, $B>0$, $C<0$, thus fulfilling the condition that the determinant with $\beta^2=0$ has two positive roots, between which all determinants are positive. 

The additional eigenvalue \eqref{lambdacases} is negative between the two roots $R_S^+$ and $R_S^-$ due to the requirement $R_S^+>R_S^-$, regardless of the choice of $M^+$ and $M^-$. This situation is illustrated in Fig. \ref{A minimal}(b). 

Finally, we have to show that the trace \eqref{trace new} is negative between $R_S^+$ and $R_S^-$. We start with $R_S^+>R_S^-$ and transform this inequality,
\begin{eqnarray} \label{tracenegative}
    \frac{-\alpha_P^+(\phi^+ - \mu^+)}{\phi^+ - \omega^{-+}}&>&\frac{-\alpha_P^-(\phi^- - \mu^-)}{q(\hat{\omega}^{-+} - \mu^-)}\\
    \frac{-\alpha_P^-(\phi^- - \mu^-)}{q(\hat{\omega}^{-+} - \mu^-)}&>&-\frac{\alpha_P^-(\phi^- - \mu^-)+ \alpha_P^+(\phi^+ - \mu^+)}{q(\hat{\omega}^{-+} - \mu^-) + (\phi^+ - \omega^{-+}) }.\quad\nonumber
\end{eqnarray}
The right-hand side in the last line is the root of the trace \eqref{trace new} and the left-hand side is the lower root of the determinant \eqref{det new}. With the choice $D<0$ we have made sure that the trace is negative between the two roots of the determinant [see Fig. \ref{A minimal}(b)]. 
Our second example is obtained by exchanging the sources and sinks in the first example. This means that $\phi^+ - \mu^+>0$, $\phi^- - \mu^-<0$, $\phi^+ - \omega^{-+}<0$, $q(\hat{\omega}^{-+} - \mu^-)>0$, $R_S^+<R_S^-$, and $-\hat{\omega}^{-+}\omega^{-+} \leq 0$. Additionally, we choose again $D<0$ and obtain by a calculation similar to \eqref{tracenegative} that the root of the trace is smaller than the smaller of the two roots of the determinant, which is now $R_S^+$. 

\subsubsection{Case V: the net source-sink flow induces negative intrapatch feedbacks in both patch types and a positive feedback loop between source and sink populations}

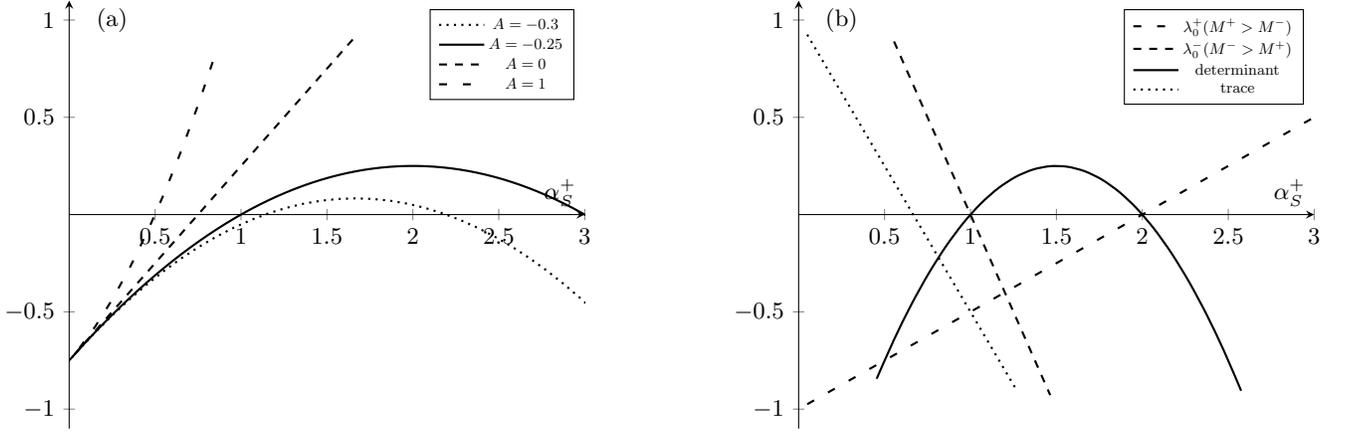
\begin{figure*}
\centering
\begin{subfigure}{}
    \centering
    \begin{tikzpicture}
    \begin{axis}[axis lines=middle,
        xmin=0, xmax=3, ymin=-1.1, ymax=1.1,
        xlabel=$\alpha_S^+$,
      restrict y to domain=-1:1,
      legend style={nodes={scale=0.65, transform shape}
        }]
     \addplot[black, samples=100, dotted, thick, unbounded coords=discard] plot (\x, -0.3*x^2+1.*x-0.75);
     \addplot[black, samples=100, smooth, thick, unbounded coords=discard] plot (\x, -0.25*x^2+1.*x-0.75);
     \addplot[black, samples=100, dashed, thick, unbounded coords=discard] plot (\x, 1.*x-0.75);
     \addplot[black, samples=100, loosely dashed, thick, unbounded coords=discard] plot (\x, 1*x^2+1.*x-0.75);
     \legend{$A=-0.3$,$A=-0.25$,$A=0$,$A=1$}
     \node[] at (axis cs: 0.25,1) {(a)};
   \end{axis}
    \end{tikzpicture}
\end{subfigure}
\hspace*{\fill}
\begin{subfigure}{}
    \centering
    \begin{tikzpicture}
    \begin{axis}[axis lines=middle,
        xmin=0, xmax=3, ymin=-1.1, ymax=1.1,
        xlabel=$\alpha_S^+$,
      restrict y to domain=-1:1,
      legend style={nodes={scale=0.65, transform shape}
        }]
     \addplot[black, samples=100, loosely dashed, thick, unbounded coords=discard] plot (\x, 0.5*x-1);
     \addplot[black, samples=100, dashed, thick, unbounded coords=discard] plot (\x, -2*x+2);
     \addplot[black, samples=100, smooth, thick, unbounded coords=discard] plot (\x, -1*x^2+3*x-2);
     \addplot[black, samples=100, dotted, thick, unbounded coords=discard] plot (\x, -1.5*x+1);
     \legend{$\lambda_0^+(M^+>M^-)$,$\lambda_0^-(M^->M^+)$,determinant,trace}
     \node[] at (axis cs: 0.25,1) {(b)};
   \end{axis} 
    \end{tikzpicture}
\end{subfigure}
 \caption{Parabolas with $Ax^2+x-0.75$, thus $B>0$, $C<0$, and with varying $A$ (a). The region between the two roots (if they exist) is positive for a fixed $A_0<0$ and all parabolas with $A>A_0$ are positive too within that region, even when $A>0$. Determinant, trace, and additional eigenvalues \eqref{lambdacases} depending on the relation of numbers of sources ($M^+$) and sinks ($M^-$) for the parameter choice of the first example with $\phi^+ - \mu^+<0$, $\phi^- - \mu^->0$, $\phi^+ - \omega^{-+}>0$, $q(\hat{\omega}^{-+} - \mu^-)<0$, $R_S^+>R_S^-$, and $\hat{\omega}^{-+}\omega^{-+} \geq 0$ (b). The trace and additional eigenvalues are always negative between the determinants roots. \label{A minimal}}
\end{figure*}

The third example is given by $\phi^+ - \mu^+>0$, $\phi^- - \mu^-<0$, $\phi^+ - \omega^{-+}<0$, $q(\hat{\omega}^{-+} - \mu^-)<0$, \ $-R_S^+<R_S^-$, and $-\hat{\omega}^{-+}\omega^{-+} > 0$, which implies $\hat{\omega}^{-+}<0$. Now an increasing net source-sink flow $\alpha_S^+$ induces a negative intrapatch feedback \eqref{intrapatch feedback S} for both types of patches. Now we find according to Fig. \ref{A minimal} that $A(\beta^2)<0$ is minimal if $\beta^2=d^+d^-$, implying
\begin{equation} \label{result intermediate}
|\hat{\omega}^{-+}| > {\mu^-} \left(\frac{\omega^{-+}}{\phi^+}-1\right)\, .
 \end{equation}
The above choice of parameters satisfies $C<0$ and $D<0$ for the expressions \eqref{C} and \eqref{D}. The condition $B>0$ for the expression \eqref{B} is also satisfied since $R_S^+<R_S^-$. Condition \eqref{B1B2C} can be satisfied without conflict with the other conditions by making $\alpha_P^-$ or $|\phi^--\mu^-|$ large enough. In this way, we will additionally achieve that $E$ is negative, such that the trace \eqref{trace new} of the reduced Jacobian is always negative. The final condition that we need to satisfy is a negative additional eigenvalue \eqref{lambdacases} between the two roots of the determinant. For $M^+<M^-$, the eigenvalue $\lambda_0^k$ is negative for all $\alpha_S^+$. For $M^+>M^-$, it becomes negative for $\alpha_S^+ > R_S^+$. Now, from Fig. \ref{A minimal}(a) we can conclude that the two roots of the determinant with $\beta^2=d^+d^-$ both are to the right-hand side of $R_S^+$. For $\beta^2=0$, we have $A>0$ and the determinant is a parabola that opens upwards and has its larger root at $R_S^+$. For $\beta^2=d^+d^-$, the curvature becomes negative, which means that the two roots of the determinant must lie to the right of $R_S^+$.
An analogous calculation can be done if $\phi^+ - \mu^+<0$, $\phi^- - \mu^->0$, and $-R_S^+>R_S^-$.

\section{Influence of the dispersal turnovers on stability \label{Appendix_DT}}

Next, we analyze the conditions under which either a sufficiently large or an intermediate dispersal turnover $\alpha_C^+$ stabilizes the system. These correspond to the cases VI and VII which are outlined in Sec. \ref{sec DT}. Proceeding similarly as before, we write the trace \eqref{trace} and the determinant \eqref{det} as functions of $\alpha_C^+$,
\begin{equation}
 \text{tr}(\alpha_C^+) = D_C \alpha_C^+ + E_C\; ,
\end{equation}
and
\begin{equation}
\label{det C}
 \text{det}(\alpha_C^+) = A_C (\beta^2) (\alpha_C^+)^2 + B_C (\beta^2) \alpha_C^+ + C_C (\beta^2).
\end{equation}
All three coefficients in the determinant depend on $\beta^2$ if the net source-sink flow $\alpha_S^+$ does not vanish.

\subsection{Case VI: Stabilization for large dispersal turnover}
\label{Appendix_LDT}

If the source-sink dynamics shall be stable for $\alpha_C^+ \to \infty$, then we require $D_C<0$ and $A_C(\beta^2)>0$ for all $\beta \in S$. Further the additional eigenvalue
\begin{widetext}
\begin{equation} \label{lambdacases C}
    \lambda_0^k = 
     \begin{cases}
       \alpha_C^+(\hat{\omega}^{-+}-\omega^{+-}) + \alpha_S^+ (\phi^+ - \omega^{-+}) + \alpha_P^+(\phi^+ - \mu^+) \quad \text{if}\;  M^+>M^- \\
       \alpha_C^+q(\hat{\omega}^{+-}-\omega^{-+}) + \alpha_S^+ q(\hat{\omega}^{-+} - \mu^-) + \alpha_P^-(\phi^- - \mu^-) \quad \text{if}\;  M^->M^+
     \end{cases}
\end{equation}
\end{widetext}
given by \eqref{lambda add} must become negative for large $\alpha_C^+$. From the first two requirements
\begin{equation}
\label{A_C(beta)}
A_C (\beta^2) = q(\hat{\omega}^{+-}-\omega^{-+})(\hat{\omega}^{-+}-\omega^{+-})(1 - \frac{\beta^2}{d^+d^-})>0
\end{equation}
and
\begin{equation}
D_C = \hat{\omega}^{+-}-\omega^{-+}+q(\hat{\omega}^{-+}-\omega^{+-})  <0
\end{equation}
we find that 
\begin{eqnarray}
\label{result alpha c large}
 \omega^{-+} &> \hat{\omega}^{+-} \nonumber\\
 \omega^{+-} &> \hat{\omega}^{-+} \, .
\end{eqnarray}
for all $\beta^2<d^+d^-$. For $\beta^2=d^+d^-$ we have $A_C=0$, and the sign of the determinant for large $\alpha_C^+$ is not determined by $A_C$  but by $B_C$. We therefore have to add the condition $B_C(d^+d^-)>0$. To evaluate this condition further, we define again the resistances of sources and sinks $R_C^+$ and $R_C^-$ as
\begin{eqnarray}
    R_C^+ &= -  \frac{\alpha_P^+(\phi^+ - \mu^+) +\alpha_S^+\phi^+}{\hat{\omega}^{+-}-\omega^{-+}} \, , \nonumber\\ \quad R_C^- &= -\frac{\alpha_P^-(\phi^- - \mu^-) -q\alpha_S^+\mu^-}{q(\hat{\omega}^{-+}-\omega^{+-})} \, .
\end{eqnarray}
These are identical to the roots of the determinant for $\beta^2=0$. Compared to the resistances to the net source-sink flow \eqref{resistance}, the denominators are replaced by the dispersal-induced intrapatch feedbacks $\alpha_C^+(\hat{\omega}^{+-}-\omega^{-+})$ in sources and $q\alpha_C^+(\hat{\omega}^{-+}-\omega^{+-})$ in sinks. 
The numerators
\begin{equation}
    \alpha_P^-(\phi^- - \mu^-)-q\alpha_S^+\mu^- \quad \text{ and } \quad \alpha_P^+(\phi^+-\mu^+)+\alpha_S^+\phi^+ 
\end{equation}
represent extended local feedbacks, which have an additional term due to source-sink dynamics depending on the net source-sink flow $\alpha_S^+$. From
\begin{eqnarray}
&B_C(d^+d^-) = (\hat{\omega}^{+-}-\omega^{-+})[\alpha_P^-(\phi^- - \mu^-)-q\alpha_S^+\mu^-] \nonumber\\
&+q(\hat{\omega}^{-+}-\omega^{+-}) [\alpha_P^+(\phi^+-\mu^+)+\alpha_S^+\phi^+]>0 \label{B_C}
\end{eqnarray}
follows that stability for large $\alpha_C^+$ is only possible if the extended local feedback is negative for at least one type of patches. If both patches have a negative extended local feedback, then the system is already stable for $\alpha_C^+=0$ according to \eqref{result local}, and thus an increase of dispersal turnover has no positive effect on stability.

More interesting is the case where the extended local feedbacks \eqref{extendedlocalfeedback} have opposite signs, since then the system is unstable for $\alpha_C^+=0$. For a large dispersal turnover to be stabilizing we find according to condition \eqref{result alpha c large} that $\alpha_C^+$ needs to induce a negative intrapatch feedbacks. Additionally from condition \eqref{B_C} follows that the absolute value of the resistance of the patch type with a negative extended local feedback has to be larger than the absolute value of the resistance of the other.

\subsection{Case VII: Stabilization for intermediate dispersal turnover}
\label{Appendix_IDT}

To find an intermediate region of stability we follow the same approach as for $\alpha_S^+$. We set $\alpha_S^+=0$ for the sake of simplicity. This means that the coefficients $B_C$ and $C_C$ are now independent of $\beta^2$. As in the previous subsection we assume that $A_C(\beta^2)<0$ for at least one $\beta \in S$, $B_C>0$, and $C_C<0$. From the expression \eqref{A_C(beta)} we can deduce that $A_C(\beta^2)<0$ is actually fulfilled for all $\beta^2 \neq d^+d^-$ if the induced intrapatch feedbacks have opposite signs, i.e., if one of the two conditions \eqref{result alpha c large} is satisfied while the other is violated. For $\beta^2 = d^+d^-$ the determinant is a linear function of $\alpha_C^+$ with a positive slope, and its root is smaller than all roots of determinants with smaller $\beta^2$ [see Fig. \ref{A minimal}(a)]. The intermediate region of stability is again bounded by the roots of the determinant with a minimal $A_C(\beta^2)$, and we therefore focus on the situation $\beta=\beta_0$. From 
\begin{equation}
\label{condition CC}
    C_C= [\alpha_P^+(\phi^+-\mu^+)][\alpha_P^-(\phi^--\mu^-)]<0
\end{equation}
we conclude that the local feedback \eqref{local feedback} of sources and sinks have opposite signs. Then the assumption $B_C>0$ requires that the induced intrapatch feedback  [$\alpha_C^+(\hat{\omega}^{+-}-\omega^{-+})$ and $q\alpha_C^+(\hat{\omega}^{-+}-\omega^{+-})$] and the local feedback \eqref{local feedback} need to have opposite signs for each patch type respectively. We further note that the roots of the determinant only exist if
\begin{equation}
\label{condition roots C}
    B_C^2-4A_C C_C>0\; .
\end{equation}
Since the resistances $R_C^+$ and $R_C^-$ are the absolute values of the determinants roots, the additional eigenvalue \eqref{lambdacases C} is negative between the roots if we further require that the resistance of the patch with a positive local feedback \eqref{local feedback} is lower than the resistance of the other patch type [see also Fig. \ref{A minimal}(b)]. Then the intermediate region of stability is bounded by the roots of the determinant with $\beta=\beta_0$ [see Fig. \ref{A minimal}(a)].

\begin{table*}
\begin{ruledtabular}
\begin{tabular}{lrrrrrrrrrr}
Fig. \ref{plot results} &$\phi_i^+$&$\mu_i^+$&$\phi_i^-$&$\mu_i^-$&$\omega_i^{-+}$&$\hat{\omega}_i^{-+}$&$\omega_i^{+-}$&$\hat{\omega}_i^{+-}$&$\alpha_{P_i}^+/\alpha_{P_i}^-$&$\alpha_{P_i}^+/\alpha_i^+$\\
\hline
 (a) and (b) & $1$ & $1/2$ & $1/2$ & $1$ & $3/2$ & $1/2$ & $0$ & $0$ & $1$ & $0.1$\\
 (c) and (d) & $0.75$ & $0.7$ & $1/4$ & $1/2$ & $1.75$ & $-1/2$ & $0$ & $0$ & $3$ & $0.1$\\
 (e) and (f) & $1/2$ & $1$ & $2$ & $1$ & $1.75$ & $-1$ & $0$ & $0$ & $1$ & $0.1$\\
 (g) and (h) & $1$ & $2$ & $2$ & $1$ & $1.75$ & $1/2$ & $2$ & $1$ & $1$ & $0.1$
\end{tabular}
\end{ruledtabular}
\caption{Parameter values of exponent parameters and turnover rates used in the plots in Figs. \ref{plot results}(a) to \ref{plot results}(h) for metapopulations [(a), (c), (e), and (g)] and metacommunities with $N=10$ [(b), (d), (f), and (h)]. The last column shows the ratio of the local turnover to the total turnover $\alpha_{P_i}^+/\alpha_i^+$, which is only relevant for the species within the respective metacommunity for which no turnover rates are varied. \label{numerical parameters}}
\end{table*}

\FloatBarrier
\section{Details of the numerical analysis}
\label{Appendix_NA}

For each set of exponent and scale parameters, we generated an ensemble of 100.000.000 biregular systems with variable patch numbers drawn uniformly from 10 to 25. Metacommunities with $N=10$ species were generated using the niche model \cite{williams2000}, in which the number of links were drawn using a $\beta$ distribution such that the mean connectance is $C=0.1$. The link strengths $l_{ij}$ were drawn from a narrow Gaussian distribution with mean value $0$ and a $10\%$ coefficient of variation. Based on these link strengths, we defined the inter-specific exponent parameters as $\phi_{ij}^k=\mu_j^k|l_{ij}|/\sum_{n=1}^N |l_{in}|$ and $\mu_{ij}^k=\phi_j^k|l_{ij}|/\sum_{n=1}^N |l_{nj}|$, such that the exponent of growth (loss) of species $i$ became proportional to the exponent of loss (growth) of species $j$ with a prey (predator) centric normalization of the link strength. Further we employ allometric scaling \cite{brown2004,otto2007} for the local turnover rates $\alpha_{P_i}^k=10^{-2n_i}$. We chose $q=1$ for simplicity. Other values of $q$ shifted the transition lines, but the qualitative results were not affected by changes in $q$.

Our parameter choices were such that the stability criteria I to VII found in the previous section can be indicated easily in the plots. All remaining parameters were fixed to the values given in TABLE \ref{numerical parameters}.

\providecommand{\noopsort}[1]{}\providecommand{\singleletter}[1]{#1}%

\end{document}